\documentclass[12pt]{article}
\usepackage{epsfig,amssymb,amsmath,psfrag}

\catcode`\@=11
\font\cmss=cmss12 
\def\1{\hbox{{1}\kern-.25em\hbox{l}}}
\def\bfZ{\relax{\hbox{\cmss Z\kern-.4em Z}}}

\def \thesection {\arabic{section}.}

\def \be  {\begin{equation}}
\def \ee  {\end{equation}}
\def \ba  {\begin{eqnarray}}
\def \ea  {\end{eqnarray}}
\def \baa {\begin{eqnarray*}}
\def \eaa {\end{eqnarray*}}
\def \bb  {\begin {thebibliography} }
\def \eb  {\end{thebibliography}}
\def \lab #1 {\label{#1}}

\newcommand\re[1]{(\ref{#1})}
\def \qqquad {\qquad\quad}

\def \matrix #1 {\left(\begin{array}{cc} #1 \end{array}\right)}

\def \tr {\mathop{\rm tr}\nolimits}

\def \e  {\mathop{\rm e}\nolimits}

\newcommand \widebar [1] {\overline{#1}}

\newcommand \vev [1] {\langle{#1}\rangle}

\newcommand{\as}{\ifmmode\alpha_{\rm s}\else{$\alpha_{\rm s}$}\fi}
\newcommand{\asbar}{\ifmmode\bar{\alpha}_{\rm s}\else{$\bar{\alpha}_{\rm s}$}\fi}

\newcommand{\ft}[2]{{\textstyle\frac{#1}{#2}}}

\font\cmss=cmss12 
\def\inbar{\,\vrule height1.5ex width.4pt depth0pt}
\def\IC{\relax\hbox{$\inbar\kern-.3em{\rm C}$}}
\def\IZ{\relax{\hbox{\cmss Z\kern-.4em Z}}}
\def\IR{{\hbox{{\rm I}\kern-.2em\hbox{\rm R}}}}

\def\IP{{\hbox{{\rm I}\kern-.2em\hbox{\rm P}}}}
\def\II{\hbox{{1}\kern-.25em\hbox{l}}}

\def\numberbysection{\@addtoreset{equation}{section}
                     \def\theequation{\thesection\arabic{equation}}}
\numberbysection

\newbox\lett\newdimen\lheight\newdimen\lwidth
\def\ontop#1#2{\setbox\lett=\hbox{#2}\lheight\ht\lett
\multiply\lheight by 12 \divide\lheight by 10\relax%
\lwidth\wd\lett \multiply\lwidth by 8 \divide\lwidth by 10\relax #2\kern-\lwidth%
\raise\lheight\hbox{{$\scriptstyle #1$}}\kern.1ex}

\psfrag{dots}[cc][cc]{$...$}
\psfrag{Z1}[cc][cc]{$\scriptstyle Z_1$}
\psfrag{Z2}[cc][cc]{$\scriptstyle Z_2$}
\psfrag{Z3}[cc][cc]{$ $}
\psfrag{ZN-1}[cc][cc]{$ $}
\psfrag{ZN}[cc][cc]{$\scriptstyle Z_L$}

\begin{document}

\begin{titlepage}
\begin{flushright}
\begin{tabular}{l}
LPT--Orsay--05--16 \\
hep-th/0503137
\end{tabular}
\end{flushright}

\vskip1cm

\centerline{\large \bf Superconformal operators in Yang-Mills theories on the
light-cone}

\vspace{1cm}

\centerline{\sc A.V. Belitsky$^a$, S.\'E. Derkachov$^{b}$,
                G.P. Korchemsky$^c$, A.N. Manashov$^{d,e}$}

\vspace{10mm}

\centerline{\it $^a$Department of Physics and Astronomy, Arizona State
University} \centerline{\it Tempe, AZ 85287-1504, USA}

\vspace{3mm}

\centerline{\it  $^b$ Department of Mathematics, St Petersburg Technology
Institute,} \centerline{\it St.-Petersburg, Russia}

\vspace{3mm}

\centerline{\it $^c$Laboratoire de Physique Th\'eorique\footnote{Unit\'e
                    Mixte de Recherche du CNRS (UMR 8627).},
                    Universit\'e de Paris XI}
\centerline{\it 91405 Orsay C\'edex, France}

\vspace{3mm}

\centerline{\it $^d$Department of Theoretical Physics,  St.-Petersburg State
University}

\centerline{\it 199034, St.-Petersburg, Russia}

\vspace{3mm}

\centerline{\it $^e $Institut f{\"u}r Theoretische Physik, Universit{\"a}t
Regensburg} \centerline{\it D-93040 Regensburg, Germany}

\def\thefootnote{\fnsymbol{footnote}}%
\vspace{1cm}

\centerline{\bf Abstract}

\vspace{5mm}

We employ the light-cone superspace formalism to develop an efficient approach
to constructing superconformal operators of twist two in Yang-Mills theories
with $\mathcal{N} = 1,2,4$ supercharges. These operators have an autonomous
scale dependence to one-loop order and determine the eigenfunctions of the
dilatation operator in the underlying gauge theory. We demonstrate that for
arbitrary $\mathcal{N}$ the superconformal operators are given by remarkably
simple, universal expressions involving the light-cone superfields. When
written in components field, they coincide with the known results obtained by
conventional techniques.

\end{titlepage}

\setcounter{footnote} 0

\thispagestyle{empty}



\newpage

\pagestyle{plain} \setcounter{page} 1

\section{Introduction}

The Operator Product Expansion (OPE) is a powerful tool in quantum field theories
\cite{Wil69} whose applications range from the second order phase transitions in
condensed matter physics to deep-inelastic scattering in QCD. It allows one to
expand a product of local operators over a set of composite Wilson operators of
increasing twist (= canonical dimension minus Lorentz spin) and express correlation
functions in terms of the structure constants of the corresponding operator algebra.
In an interacting theory, the Wilson operators  mix under renormalization and
acquire nontrivial anomalous dimensions. Construction of the Wilson operators
possessing an autonomous scale dependence and finding the spectrum of their
anomalous dimensions both at weak and strong coupling is an important problem in
four-dimensional gauge theories. At weak coupling, it can be solved by calculating
the corresponding mixing matrices order by order in the coupling constant and
diagonalizing them perturbatively.  At strong coupling, the problem awaits its
solution. The maximally supersymmetric $\mathcal{N} = 4$ super-Yang-Mills (SYM)
theory takes an exceptional place among all gauge theories from this standpoint,
since the gauge/string correspondence---a strong/weak duality \cite{Mal97} between
gauge and string theories---allows one to map Wilson operators into certain states
of the string on the $AdS_5\times S^5$ background and identify the anomalous
dimension of the former as the energy of the latter (for a comprehensive review,
see \cite{Tse04}).

In this paper we address the problem of constructing Wilson operators having an
autonomous scale dependence in (super) Yang-Mills theory with an arbitrary number
of supercharges $0 \le \mathcal{N}\le 4$. We restrict the analysis to the leading,
twist-two operators. These operators possess a two-particle structure and can be
obtained by expanding a nonlocal light-cone operator in the Taylor series,
symbolically
\be
\phi_1(z_1n_\mu) \phi_2(z_2n_\mu) = \sum_{k\ge 0} \frac{z_{21}^k}{k!}\, {O}_k(z_1
n_\mu)\, ,
\label{phi-phi}
\ee
where $O_k(0) =  \phi_1(0) \partial_+^k \phi_2(0)$ is a local composite operator,
$z_{21}\equiv z_2-z_1$ and $\partial_+ \equiv n \cdot \partial$ is the light-cone
derivative. Here $n_\mu$ is a light-like vector such that $n_\mu^2 = 0$ and the
scalar variables $z_j$ (with $j = 1,2$) define the position of a generic field
$\phi_j(z_j n_\mu)$ on the light-cone. In Eq.~\re{phi-phi}, it is tacitly assumed
that the gauge invariance is restored by inserting a gauge link between the two
fields in the left-hand side and replacing the ordinary derivatives by the covariant
ones in the right-hand side. The operators $O_k(0)$ mix under renormalization with
operators of the same canonical dimension involving total derivatives $\partial^\ell_+
O_{k-\ell}(0)$ as well as with other twist-two operators having a different
``partonic'' content.

It is well-known that to one-loop order the form of multiplicatively
renormalizable operators is constrained by conformal invariance of the classical
(super) Yang-Mills theory (for a review, see \cite{BraKorMul03}). Although the
conformal symmetry is broken on the quantum level (except for $\mathcal{N}=4$
SYM) this affects the mixing matrix of Wilson operators starting from two loops
only \cite{BelMul98}. As a result, the one-loop mixing matrix inherits the
symmetry of the classical Lagrangian and its eigenstates $\mathcal{O}_N(0)$ can
be classified according to representations of the (super)conformal $SU(2,2 |
\mathcal{N})$ group and, more precisely, to its collinear subgroup $SL(2|
\mathcal{N})$ acting on the fields ``living'' on the light-cone.

For instance, in the $\mathcal{N}=0$ theory (that is, pure gluodynamics), the twist-two
(parity-even) operators are built from gauge fields only and have the form \cite{Mak81}
\be
\mathcal{O}_N(x) = (i\partial_+)^N \tr \left[F_{+\mu} (x)\,
\mathrm{C}_N^{5/2}
\left( \frac{ \stackrel{\rightarrow}{D}_+ -
\stackrel{\leftarrow}{D}_+ }{ \stackrel{\rightarrow}{\partial}_+ +
\stackrel{\leftarrow}{\partial}_+} \right) F_{+}^{~\mu}(x)\right],
\label{GG}
\ee
where $\mathrm{C}_N^{5/2}(\xi)$ is the Gegenbauer polynomial and $\partial_+ =
\stackrel{\rightarrow}{\partial}_+ + \stackrel{\leftarrow}{\partial}_+$ is the
total derivative. Also, the arrows indicate the fields to which the derivatives
are applied. The subscripts `+' on symbols exhibit the projection of the
corresponding Lorentz indices onto the light-cone, e.g., for the fields
strength $F_{+\mu}(x) = n^\nu F_{\nu\mu}^a(x)t^a$, and the covariant derivative,
${D}_+ = n^\mu {D}_\mu$.  The composite operator \re{GG} is transformed under
the conformal $SL(2)$ transformations as a primary field with the conformal
spin $j = N + 3$. It can be written as $\mathcal{O}_N(x) \sim {O}_N +
\sum_{k\ge 1} c_k \partial_+^k {O}_{N-k}$ with ${O}_{N} =
F_{+\mu}^a(x)\partial_+^N F_{+\mu}^a(x)$ and the coefficients $c_k$ uniquely
fixed by the conformal symmetry. Conformal invariance ensures that the operator
\re{GG} diagonalizes the one-loop dilatation operator of the $\mathcal{N}=0$
theory and, therefore, $\mathcal{O}_N(x)$ has an autonomous scale dependence.
The corresponding anomalous dimension depends on the conformal $SL(2)$ spin
of the operator \re{GG}, but the explicit form of this dependence is not
fixed by the conformal symmetry.

In supersymmetric Yang-Mills theories with $\mathcal{N} \ge 1$ supercharges, the
conformal operators \re{GG} do not have an autonomous scale dependence since they
mix under renormalization with similar twist-two operators of the same conformal
spin but built from the gauge field super-partners---gauginos and scalars. It
is a linear combination of the $SL(2)$ conformal operators that diagonalizes the
mixing matrix and has the property of multiplicative renormalizability. To one-loop
order, the resulting twist-two operators belong to an irreducible representation
of the $SL(2|\mathcal{N})$ group and carry a definite value of the superconformal
spin. We shall refer to them as superconformal operators. As before, the $SL(2|
\mathcal{N})$ invariance allows one to determine the explicit form of these
operators without going through diagrammatical calculation of the mixing matrix
but it does not fix the dependence of their anomalous dimensions on the superconformal
spin.

One approach to constructing the superconformal operators in the SYM theory proposed
in Ref.~\cite{BukFroLipKur85} consists in examining the properties of conformal
operators, like \re{GG}, under supersymmetric transformations belonging to the $SL(2|
\mathcal{N})$ group. Repeatedly applying supersymmetric variations to \re{GG},
one can construct a supermultiplet of twist-two operators defining an irreducible
representation of the $SL(2|\mathcal{N})$ group. The operators entering the
supermultiplet diagonalize the one-loop dilatation operator of SYM theory and
have the same anomalous dimension. Although the above procedure is straightforward,
its implementation in SYM theory with $\mathcal{N}> 1$ supercharges becomes
extremely cumbersome due to a large number of operators involved and growing size
of supermultiplets. The resulting expressions for the superconformal operators
look differently for different $\mathcal{N}$ and do not exhibit any universal
structure.

In this paper we propose another approach to constructing the superconformal
operators which allows one to treat simultaneously SYM theories with an arbitrary
number of supercharges $\mathcal{N}$. It relies on the light-cone formulation of
SYM theory \cite{BriLinNil83,Man83} and takes full advantage of the $SL(2|\mathcal{N})$
superconformal group. A detailed account on this formulation can be found in Ref.\
\cite{BelDerKorMan04} and we summarize here its main features. Quantizing the gauge
theory in the light-cone gauge $n \cdot A^a(x) = A^a_+(x)=0$, one can use the
equations of motion in order to eliminate dynamically dependent fields and
reformulate the SYM theory on the light-cone in terms of propagating fields only.
In case of the maximally supersymmetric, $\mathcal{N} = 4$ SYM they include transverse
components of the gauge field $A_\perp(x)=(A(x),\bar A(x))$ of helicity $\pm 1$, eight
gaugino fields $\lambda^A(x)$ and $\bar\lambda_A(x)$ of helicity $\pm \frac12$ and
six scalars $\bar\phi_{AB}(x)$ (with ${\scriptstyle A,B}=1,...,4$). Introducing the
fermionic coordinates $\theta^A$ one can rewrite the Lagrangian of the theory in
terms of two distinct chiral superfields ${\Phi}(x^\mu,\theta^A)$ and ${\Psi}(x^\mu,
\theta^A)$ (with ${\scriptstyle A}=1,\ldots,\mathcal{N}$) which comprise all
dynamically independent propagating fields. The latter are the coefficients in the
Taylor expansion of the superfields in powers of the Grassmann (or odd) coordinates
$\theta^A$. The explicit expressions for the chiral superfields
are
\be
{\Phi} (x) = \partial_+^{-1} A(x)\,,\qquad {\Psi} (x) = - \partial_+
\bar A(x)
\label{M=0-field}
\ee
for $\mathcal{N}=0$,
\ba\label{M=1-field}
{\Phi}  (x ,  \theta ) &=&
\partial_+^{-1}A(x) + \theta\, \partial_+^{-1} \bar\lambda  (x)
\,,\qquad {\Psi}  (x, \theta) = - \lambda(x) + \theta\partial_+ \bar A(x)
\ea
for $\mathcal{N}=1$,
\ba
{\Phi} (x , \theta^A ) &=&
\partial_+^{-1} A(x)
+ \theta^A \partial_+^{-1}\bar\lambda_A (x) + \frac{i}{2!} \varepsilon_{AB}
\theta^A \theta^B \bar \phi (x)  \, ,
\nonumber\\
{\Psi} (x,\theta^A) &=& i \phi(x) - \varepsilon_{AB} \theta^A \lambda^B (x) +
\frac1{2!} \varepsilon_{AB} \theta^A\theta^B \partial_+
\bar A (x) \, ,
\label{M=2-field}
\ea
for $\mathcal{N}=2$, and
\begin{eqnarray}
{\Phi} (x, \theta^A) &=& {\Psi} (x, \theta^A) =
\partial_+^{-1}A(x) +\theta^A
\partial_+^{-1}\bar\lambda_A (x) + \frac{i}{2!} \theta^A \theta^B \bar \phi_{AB}
(x)
\nonumber\\
&+&\!\!\! \frac{1}{3!} \varepsilon_{ABCD} \theta^A \theta^B \theta^C \lambda^D
(x) - \frac{1}{4!} \varepsilon_{ABCD} \theta^A \theta^B \theta^C \theta^D
\partial_+ \bar{A} (x)
\label{M=4-field}
\end{eqnarray}
for $\mathcal{N}=4$. In the latter case, the two superfields coincide since each
of them comprises all propagating modes. This is contrasted to $0 \le \mathcal{N}
\le 2$ models where only half of the propagating fields can be accommodated into
either of the two superfields.

Similar to \re{phi-phi}, all possible twist-two operators in the
$\mathcal{N}-$extended SYM theories can be obtained by expanding the $\Psi\Psi-$,
$\Phi\Phi-$ and $\Phi\Psi-$products of two superfields located on the light-cone
$x^\mu = z n^\mu$. Let us define $Z=(z,\theta^A)$ (with ${\scriptstyle A} = 1,
\ldots, \mathcal{N}$) as a point in the $(\mathcal{N}+1)-$dimensional light-cone
superspace and use a shorthand notation $\mathbb{O}(Z_1,Z_2)$ for the
superfield bilinears $\tr \left[\Psi(Z_1) \Psi(Z_2)\right]$, $\tr \left[\Phi(Z_1)
\Phi(Z_2)\right]$ and $\tr \left[\Phi(Z_1) \Psi(Z_2)\right]$. Then, the generating
function for twist-two operators in $\mathcal{N}-$extended SYM theories looks like
\be
\label{O-def}
\mathbb{O}(Z_1,Z_2) = \sum_{N\ge 0} \frac{z_{21}^N}{N!} \sum_{n,m=0}^\mathcal{N}
\theta_{1}^{A_{1}}\!\ldots \theta_1^{A_{n}} \theta_{2}^{B_{1}}\!\ldots
\theta_2^{B_{m}} \, \mathcal{O}_{N;A_1,\ldots,A_n;B_1,\ldots,B_n}(z_1 n_\mu) \,.
\ee
Here the expansion in the right-hand side involves all possible powers of odd
variables $\theta_{1,2}^{A}$ and the expansion coefficients define the composite
twist-two operators $\mathcal{O}_{N,\{A\},\{B\}}(x)$.

The twist-two operators in the right-hand side of \re{O-def} mix under
renormalization and their anomalous dimensions can be extracted from the
renormalization group equation for the nonlocal light-cone superfield
operator $\mathbb{O} (Z_1,Z_2)$ \cite{BelDerKorMan04}
\be
\left\{ \mu\frac{\partial}{\partial\mu} +
\beta_\mathcal{N}(g)\frac{\partial}{\partial g}
+ 2 \gamma_\mathcal{N}(g) \right\} \mathbb{O}(Z_1,Z_2) = - \frac{g^2 N_c}{8\pi^2}
\left[ \mathbb{H}\cdot \mathbb{O}\right](Z_1,Z_2) + \mathcal{O}(g^4)\, .
\label{EQ}
\ee
Here $\beta_\mathcal{N}(g)$ is the beta-function in the SYM theory and
$\gamma_\mathcal{N}(g)$ is the anomalous dimension of the superfields. In the
light-like axial gauge $A_+(x)=0$ one has $\gamma_\mathcal{N}(g)=\beta_\mathcal{N}(g)/g$.
The integral operator $\mathbb{H}$ defines a representation of the one-loop dilatation
operator of SYM theory on the space spanned by the operators $\mathbb{O}(Z_1,Z_2)$.
Its explicit form has been found in Ref.~\cite{BelDerKorMan04} for arbitrary number
of supercharges $0\le \mathcal{N} \le 4$. To solve the evolution equation \re{EQ}
it suffices to solve the spectral problem for the operator $\mathbb{H}$. Its
eigenvalues, $E (j)$, provide the spectrum of one-loop anomalous dimensions of
multiplicatively renormalizable Wilson operators in the $\Psi\Psi-$, $\Phi\Phi-$
and $\Phi\Psi-$sectors, while the corresponding eigenstates determine their explicit
form.

It turns out that the anomalous dimensions of twist-two operators
$\mathcal{O}_n(0)$ in $\mathcal{N}-$extended SYM theories have a universal
form~\cite{BelDerKorMan04}
\be\label{gg}
\gamma = \frac{g^2 N_c}{8\pi^2}\left(E + 2
\gamma_\mathcal{N}^{\scriptscriptstyle (0)} \right) + \mathcal{O}(g^4)\,,
\ee
where $\gamma_\mathcal{N}^{\scriptscriptstyle (0)}$ is the one-loop correction to
the anomalous dimension of the superfield,
$\gamma_\mathcal{N}^{\scriptscriptstyle (0)}= - {11}/{6}, - 3/2 , - 1, 0$ for
$\mathcal{N}=0,1,2,4$, respectively, and $E$ is the eigenvalue of the one-loop
dilatation operator $\mathbb{H}$. Depending on the sector to which the Wilson
operator belongs it is given by
\begin{itemize}
\item  $\Psi\Psi-$ and $\Phi\Phi-$sectors
\be
E_{\Psi\Psi}(j) = E_{\Phi\Phi}(j) = 2\left[\psi(j)-\psi(1)\right] \, ,
\label{H-PsiPsi-J}
\ee
\item  $\Phi\Psi-$sector
\ba\label{H-PhiPsi-J}
E_{\Psi\Phi}(j) =
\psi\left(j + 2 - \mathcal{N}/2\right)
\!\!\!&+&\!\!\!
\psi\left(j - 2 + \mathcal{N}/2\right)
- 2 \psi(1) \\
&+&\!\!\!
(-1)^{j+\mathcal{N}/2}
\frac{\Gamma(j-2+\mathcal{N}/2)}{\Gamma(j+2-\mathcal{N}/2)}\Gamma(4-\mathcal{N})
\, , \nonumber
\ea
\end{itemize}
where $j$ is the superconformal $SL(2|\mathcal{N})$ spin of the corresponding Wilson
operator and $\psi(x)=d\ln\Gamma(x)/dx$ is the Euler dilogarithm. In the $\mathcal{N}
= 4$ SYM \emph{all} twist-two operators belong to the $\Phi\Phi-$sector and their
anomalous dimension is given by \re{H-PsiPsi-J}. The corresponding Wilson operators
have been constructed in Ref.~\cite{BelDerKorMan04} through the diagonalization of
the one-loop mixing matrix. They are given by linear combinations of two-particle
operators built from propagating fields $\phi_k (x) = \{\partial_+A, \partial_+\bar A,
\lambda^A,\bar\lambda_A,\bar \phi_{AB}\}$ and have the following general form
\be
\label{P-ab}
\mathcal{O}_n(0) = \sum_{k,j} \phi_k(0)\,
\mathbb{P}_n^{(k,j)}(\stackrel{\rightarrow}{\partial}_+ ,
\stackrel{\leftarrow}{\partial}_+)\, \phi_j(0) \, ,
\ee
where $\mathbb{P}_n^{(k,j)}(x_1,x_2)$ are some polynomials in light-cone derivatives
with arrows indicating, as usual, where they are applied.

In the light-cone formalism, the fields $\phi_k (x)$ are the components of the
light-cone superfields, Eqs.~\re{M=0-field} -- \re{M=4-field}. This suggests
that the Wilson operators \re{P-ab} can be constructed as
\be
\label{SuperConfOper}
\mathcal{O}_n(0) =
P_n \left( \partial_{Z_1}; \partial_{Z_2} \right)
\mathbb{O}(Z_1,Z_2) \bigg|_{Z_1=Z_2=0}
,
\ee
where $\mathbb{O}(Z_1,Z_2)$ is given by the product of two superfields, Eq.~\re{O-def},
and $P_n( \partial_{Z_1}; \partial_{Z_2})$ is a polynomial in superspace derivatives
$\partial_{Z} = (\partial_{z}, \partial_{\theta^A})$, cf.\ Refs.\ \cite{DolOsb00,Kir04}.
Equation \re {SuperConfOper} generalizes the expression for the conformal operators in
the $\mathcal{N}=0$ theory, Eq.~\re{GG}, in which case the polynomial $P_n(x_1;x_2)$,
depending only on the bosonic variables, is given in terms of the Gegenbauer
polynomials
\be
P_n(x_1;x_2)\bigg|_{\mathcal{N}=0}=(x_1+x_2)^n
\mathrm{C}_n^{5/2}\left(\frac{x_1-x_2}{x_1+x_2}\right)
\,.
\ee
In SYM theories with $\mathcal{N}\ge 1$ supercharges, the polynomial $P_n (X_1;
X_2)$---depending on two bosonic and $2 \mathcal{N}$ fermionic variables via $X_i
= (x_i, \vartheta_{i, A_i})$---can be expanded in powers of Grassmann $\vartheta_{1,
A_1}$ and $\vartheta_{2, A_2}$ so that the corresponding coefficients are given by
the polynomials $\mathbb{P}_n^{(k,j)} (x_1,x_2)$, Eq.~\re{P-ab}. We shall argue that
the explicit form of these polynomials is uniquely determined by the superconformal
$SL(2|\mathcal{N})$ symmetry. Namely, the superfields $\Phi(Z)$ and $\Psi(Z)$ are
transformed under superconformal $SL(2|\mathcal{N})$ transformations according to
irreducible representations of the $SL(2|\mathcal{N})$ group specified by their
superconformal spins $j_\Phi=-1/2$ and $j_\Psi=(3-\mathcal{N})/2$, respectively. As
a result, the nonlocal light-cone operator $\mathbb{O}(Z_1,Z_2)$, Eq.~\re{O-def},
belongs to the tensor product of two $SL(2|\mathcal{N})$ representations. The
polynomial $P_n(X_1; X_2)$ is fixed by the condition that the operator $P_n
\left(\partial_{Z_1};\,\partial_{Z_2}\right)$ has to project this tensor product
onto one of its irreducible components. Then, the mixing of the local twist-two
operators \re{SuperConfOper} with other twist-two operators is protected (to one-loop
order at least) by the superconformal symmetry.

The outline of the paper is the following. In Sect.~2 we illustrate the formalism
of constructing the superconformal operators on a simple example of conventional
$SL(2)$ operators. We briefly review representations of the underlying collinear
conformal group and establish the relation between the lowest weights in the
tensor product of two $SL(2)$ modules and the twist-two conformal polynomials in
the $\mathcal{N}=0$ theory. In Sect.~3 we employ the light-cone superspace
formalism and extend our consideration to supersymmetric Yang-Mills theories. We
demonstrate that the above mentioned relation between the superconformal
operators and the lowest weights of the $SL(2| \mathcal{N})$ group is universal
and it holds true for an arbitrary number of supercharges $\mathcal{N}$. Next, in
Sect.~4 we use this construction to work out the explicit expressions for the
superconformal operators in the $\mathcal{N}=1$, $\mathcal{N}=2$ and
$\mathcal{N}=4$ SYM theories. Sect.~5 contains concluding remarks. Two appendices
give details on Jacobi polynomials and the $SL(2|\mathcal{N})$ superconformal
transformations.

\section{Conformal operators}

To illustrate our approach, we first revisit the construction of the $SL(2)$
conformal operators, Eq.~\re{GG}. We recall that twist-two operators
$\mathcal{O}_{\mu_1.....\mu_n}(0)$ are composite gauge-invariant operators built
from  covariant derivatives $D_\mu$ and fundamental fields in (super) Yang-Mills
theory $\phi_k(x) = \phi_k^a(x) t^a$, that are symmetric and traceless in
any pair of Lorentz indices. They possess the Lorentz spin $N$ and the canonical
dimension $2 + n$. It is convenient to contract all Lorentz indices with a
light-like vector $n_\mu$ (such that $n^2=0$) and introduce
\be\label{def-1}
{\cal O}_n (0)
\equiv
n^{\mu_1}\ldots n^{\mu_n}\mathcal{O}_{\mu_1\ldots\mu_n}(0)
=
\sum_{k_1 + k_2 = n} c_{k_1k_2}
\tr \left[ D_+^{k_1} \phi_1(0) D_+^{k_2} \phi_2(0) \right]
\, .
\ee
Here the local composite operators $\tr\left[ D_+^{k_1}\phi_1(0) D_+^{k_2}\phi_2(0)
\right]$ define a basis in the space of twist-two operators. Throughout the paper
we adopt the light-cone gauge $n\cdot A(x) = A_+(x) = 0$ so that all covariant
derivatives are reduced to the ordinary ones, $D_+ \equiv n \cdot D = \partial_+$.
The elementary fields entering \re{def-1} are $\phi_k = \{\partial_+ A,\partial_+
\bar A,\lambda^A,\bar\lambda_A,\phi^{AB}\}$.

The twist-two operators \re{def-1} are uniquely specified by the set of
coefficients $c_{k_1k_2}$. They are fixed from the condition that the operators
\re{def-1} enjoy an autonomous scale dependence, to one-loop order at least,
\be\label{RG}
\mu\frac{d}{d\mu} {\cal O}_n(0)
=
-\frac{g^2 N_c}{8\pi^2}
 \gamma^{\scriptscriptstyle (0)} (n) {\cal O}_n(0)
\, ,
\ee
with $\gamma^{\scriptscriptstyle (0)} (n)$ being the one-loop anomalous dimension
of the Wilson operator. Here we assumed for simplicity that the set of the twist-two
operators $\tr\left[ D_+^{k_1}\phi_1(0)D_+^{k_2}\phi_2(0) \right]$ is closed under
renormalization.

\subsection{Conformal polynomials}

Let us put into correspondence to the operator \re{def-1} the following homogeneous
polynomial in two variables
\be
\label{P} P_n (x_1,x_2)=\sum_{{k_1+k_2 = n}\atop k_1,k_2 \ge 0}
c_{k_1k_2}\,x_1^{k_1}x_2^{k_2}\,.
\ee
Then, the local operator ${\cal O}_n (0)$ can be obtained from $\mathbb{O}(z_1,z_2)
= \tr\left[\phi_1(z_1 n_\mu)\phi_2(z_2n_\mu)\right]$ as a projection
\be\label{OP}
{\cal O}_n(0)=P_n(\partial_{z_1},
\partial_{z_2})\,\mathbb{O}(z_1,z_2)\bigl|_{z_1=z_2=0}
\, ,
\ee
which establishes a correspondence between local and nonlocal operators.

To determine the polynomials $P_n(x_1,x_2)$ we require that the operators \re{OP}
have to satisfy \re{RG}. In general, for a given $n$ one expects to find a few
such operators. To enumerate them we introduce a superscript $(\ell)$ and denote
the corresponding operators and polynomials as ${\cal O}^{(\ell)}_n(0)$ and
$P_n^{(\ell)} (x_1,x_2)$, respectively. Then, assuming that the set of the
twist-two operators ${\cal O}^{(\ell)}_n(0)$ is complete, the relation \re{OP}
can be inverted as
\be
\label{Q}
\mathbb{O}(z_1,z_2)
=
\sum_{n,\ell} \Psi^{(\ell)}_n(z_1,z_2)\,{\cal O}^{(\ell)}_n(0)
\,,
\ee
where $\Psi^{(\ell)}_n(z_1,z_2)$ is yet another set of homogeneous polynomials of
degree $n$. Substituting \re{Q} into \re{OP} one finds that the two sets of
polynomials are related to each other as
\be\label{ort-1}
P_n^{(\ell)} (\partial_{z_1},\partial_{z_2}) \, \Psi_m^{(\ell')} (z_1,z_2)
\bigl|_{z_1=z_2=0} =\delta_{nm} \delta_{\ell \ell'} \, .
\ee
The polynomials $\Psi_m^{(\ell')}(z_1,z_2)$ admit a power series representation
similar to \re{P} with the expansion coefficients related to $c_{k_1k_2}$ through
the orthogonality condition \re{ort-1}.

The Wilson operators \re{def-1} are uniquely determined by the polynomials
$P_n^{(\ell)}({x_1},{x_2})$ and $\Psi_n^{(\ell)}(z_1,z_2)$. The former polynomial
projects the local twist-two operator out from the nonlocal light-cone operator,
Eq.~\re{OP}, while the latter defines the coefficient function of
$\mathcal{O}_n^{(\ell)}$ in the OPE expansion of the nonlocal light-cone
operator, Eq.~\re{Q}. It is easy to demonstrate that the renormalization group
equation for the local operators \re{RG} leads to a Schr\"odinger-like equation
for the polynomials $\Psi_n^{(\ell)}(z_1,z_2)$. To one-loop order, the nonlocal
operator $\mathbb{O}(z_1,z_2)$ satisfies the Callan-Symanzik equation
\re{EQ}
\be\label{H}
\mu\frac{d}{d\mu}\mathbb{O}(z_1,z_2)= - \frac{g^2 N_c}{8\pi^2} \left[ (
\mathbb{H} + 2 \gamma_\mathcal{N}^{\scriptscriptstyle (0)} ) \cdot \mathbb{O}
\right] (z_1, z_2) + {\cal O}(g^4) \, ,
\ee
with $\mathbb{H}$ being an integral operator acting on the light-cone coordinates
of fields \cite{AZ78,BelDerKorMan04}. Substituting \re{Q} into \re{H} and taking
into account \re{RG} one obtains
\be\label{Sch}
\left[ \mathbb{H} \cdot \Psi_n\right] (z_1,z_2) = E (n) \Psi_n (z_1,z_2) \, ,
\ee
with $\gamma^{\scriptscriptstyle (0)} (n) = E (n) + 2
\gamma_\mathcal{N}^{\scriptscriptstyle (0)}$. As was already mentioned, the
one-loop evolution operator $\mathbb{H}$ inherits the conformal symmetry of the
classical Lagrangian of the gauge theory and, therefore, its eigenfunctions
$\Psi_n (z_1,z_2)$ can be classified according to representations of the $SL(2)$
conformal group.

\subsection{$SL(2)$ conformal symmetry}

It is well-known that for nonlocal light-cone operators $\mathbb{O}(z_1,z_2)
= \tr\left[\phi_1(z_1 n_\mu)\phi_2(z_2 n_\mu)\right]$ the full $SO(4,2)$ conformal
group is reduced to its ``collinear'' $SL(2)$ subgroup. The latter acts on the
light-ray $x_\mu  = z n_\mu$ as follows,
\be\label{SL2-trans}
z \to z'=\frac{az+b}{cz+d}\,,\qquad \phi_k(z n_\mu) \to \phi_k'(z n_\mu)=
(cz+d)^{-2j_k} \phi_k\left(\frac{az+b}{cz+d}\,n_\mu\right),
\ee
with $ad-bc=1$ and $j_k$ being the conformal spin of the field $\phi_k( z n_\mu)$.
For the gauge fields ($\partial_+ A,\partial_+ \bar A$), gauginos ($\lambda^A,
\bar\lambda_A$) and scalars ($\phi^{AB}$) it takes the following values
\be\label{spins}
j_{\rm gauge} = \frac 32
\, , \qquad
j_{\rm gaugino} = 1 \, , \qquad
j_{\rm scalar} = \frac12
\, .
\ee
The field $\phi_k( z n_\mu)$ defines an $SL(2)$ representation that we shall
denote as $\mathbb{V}_{j_k}$. The representation space is spanned by polynomials
$\{1,z_k,z_k^2,\ldots\}\in \mathbb{V}_{j_k}$ which define the coefficient
functions in the expansion of the field $\phi_k( z_k n_\mu)$ around the origin
\be\label{X-Taylor}
\phi_k( z_k n_\mu) = \phi_k(0) + z_k \,\partial_+ \phi_k(0) + \frac{z_k^2}2
\,\partial_+^2 \phi_k(0) + \ldots\,.
\ee
The generators of the $SL(2)$ transformations \re{SL2-trans} can be realized as
differential operators acting on the light-cone coordinates of the fields
$\phi_k( z_k n_\mu)$
\be\label{SS}
L_{+{(k)}}=z_k^2\partial_{z_k}+2j_k z_k\,,\qquad
L_{-{(k)}}=-\partial_{z_k}\,,\qquad L_{0{(k)}}=z_k\partial_{z_k}+j_k\,,
\ee
with $j_k$ defined in \re{spins}.

The conformal invariance implies that the evolution operator $\mathbb{H}$
commutes with the two-particle $SL(2)$ generators, $[ \mathbb{H}, L_\alpha ] =0$
with $\alpha=\pm,0$ and $L_\alpha=L_{\alpha{(1)}} +L_{\alpha{(2)}}$. Therefore,
$\mathbb{H}$ is a function of the two-particle Casimir operator
\be\label{SL2-Casimir}
L^2=L_0^2+(L_+L_-+L_-L_+)/2
\ee
and its eigenstates belong to the irreducible components in the tensor product of
two $SL(2)$ modules
\be\label{SL2-dec}
\mathbb{V}_{j_1}\otimes\mathbb{V}_{j_2} =  \sum_{n\ge 0}
\mathbb{V}_{j_1+j_2+n} \, ,
\ee
with the space $\mathbb{V}_{j_1+j_2+n}$ spanned by the states
\ba\label{SL2-0}
\Psi_n^{(0)} (z_1,z_2) &=& (z_1 - z_2)^n\,,\qquad
\\[2mm]
\Psi_n^{(\ell)} (z_1,z_2) &=& (L_+)^\ell\Psi^{(0)}_n (z_1,z_2)\,, \nonumber
\ea
with $\ell\ge 1$. Here $\Psi_n^{(0)} (z_1,z_2)$ is the lowest weight, $L_-
\Psi_n^{(0)} (z_1,z_2) = 0$, and $\Psi_n^{(\ell\,)} (z_1,z_2)$ being homogeneous
polynomials in $z_1$ and $z_2$ of degree $n+\ell$. The polynomials \re{SL2-0}
diagonalize the Casimir operator
\be\label{L2-SL2}
L^2 \,\Psi_n^{(\ell)} (z_1,z_2) = (n+j_1+j_2)(n+j_1+j_2-1) \Psi_n^{(\ell)}
(z_1,z_2)
\ee
and define the eigenstates of the evolution kernel $\mathbb{H}$, Eq.~\re{Sch}.
Its eigenvalues do not depend on $\ell$ and determine the one-loop anomalous
dimension $\gamma^{\scriptscriptstyle (0)} (n)$.

Let us return to \re{Q} and substitute the polynomials $\Psi_n^{(\ell)} (z_1,z_2)$
by their expressions \re{SL2-0}. By construction, the corresponding operators
${\cal O}^{(\ell)}_n(0)$ are multiplicatively renormalizable to one-loop order
and satisfy the evolution equation \re{RG}. To determine their explicit form
we shall make use of the fact that the polynomials $\Psi_n^{(\ell)} (z_1,z_2)$,
Eq.~\re{SL2-0}, are orthogonal with respect to the $SL(2)$ invariant scalar product.

Using the isomorphism $SL(2) \sim SU(1,1)$, the scalar product between two
``single particle'' polynomials $\psi(z)$ and $\varphi(z)$ belonging to
$\mathbb{V}_j$ can be defined as~\cite{GGV-book}
\be\label{sc}
\vev{\psi|\varphi} = \int  [\mathcal{D}z]_j \, \widebar{\psi(z)}\,\varphi(z)\,,
\ee
where the integration goes over a unit disk in the complex $z-$plane with the
measure%
\footnote{Our definition of the integration measure differs from the conventional
one by the factor $1/\Gamma(2j)$ which was introduced for the later convenience
(see Sect.~4.2). }
\be
\label{SL2-measure}
\int[\mathcal{D}z]_j= \frac1{\pi\Gamma(2j-1)}\int_{|z|\le 1}
{d^2z\,(1-|z|^2)^{2j-2}}  \,.
\ee
The scalar product \re{sc} is invariant under the $SL(2)$ transformations
\re{SL2-trans} with $c=b^*$ and $d=a^*$ so that $|a|^2-|b|^2=1$. A simple
calculation shows that
\be\label{SL2-norm}
\vev{z^k | z^n} = \frac{\delta_{kn} k!}{\Gamma(2j+k)}
\,.
\ee
For states belonging to the tensor product \re{SL2-dec} the scalar product is
defined as
\be\label{sc-2}
\vev{\Psi|\Phi} = \int  [\mathcal{D}z_1]_{j_1} \int [\mathcal{D}z_2]_{j_2}\,
\widebar{\Psi(z_1,z_2)}\, \Phi(z_1,z_2)\,.
\ee
It is straightforward to verify that the $SL(2)$ generators \re{SS} satisfy the
relations  $L_0^\dagger=L_0$ and $(L_-)^\dagger=-L_+$, so that the Casimir
operator $L^2$, Eq.~\re{SL2-Casimir}, is a self-adjoint operator with respect to
the scalar product \re{sc-2}. Together with \re{L2-SL2} this property ensures
that the states \re{SL2-0} satisfy the orthogonality condition
\be\label{SL2-ortho}
\vev{\Psi_n^{(\ell)}(z_1,z_2)|\Psi_m^{(\ell')}(z_1,z_2)} \sim \delta_{nm}
\delta_{\ell\ell'}
\,.
\ee
Let us now return to \re{Q} and evaluate the scalar product of both sides of
\re{Q} with the same vector $\Psi_n^{(\ell)}(z_1,z_2)$.

Taking into account \re{SL2-ortho}, one gets the following expression for the
$SL(2)$ conformal operator
\be\label{N=0-v}
\mathcal{O}^{(\ell)}_n(0) \sim  \vev{\Psi_n^{(\ell)}(z_1,z_2)|
\mathbb{O}(z_1,z_2)} = \int  [\mathcal{D}z_1]_{j_1} \int [\mathcal{D}z_2]_{j_2}\,
\widebar{\Psi_n^{(\ell)}(z_1,z_2)}\,\mathbb{O}(z_1,z_2)\,.
\ee
Substituting
$\mathbb{O}(z_1,z_2)=\e^{z_1\partial_{w_1}+z_2\partial_{w_2}}\mathbb{O}(w_1,w_2)
\big|_{w_{1,2}=0}$, one can bring this relation to the form \re{OP} with the
$P-$polynomial determined by the scalar product
\be\label{SL2-P}
P_n^{(\ell)}(x_1,x_2) = \vev{\Psi_n^{(\ell)}(z_1,z_2)|\e^{z_1{x_1}+z_2{x_2}}}=
\vev{(L_+)^\ell\Psi_n^{(0)}(z_1,z_2)|\e^{z_1{x_1}+z_2{x_2}}}\,.
\ee
Since $(L_+)^\dagger = - L_-=\partial_{z_1}+\partial_{z_2}$, the polynomials with
$\ell\ge 1$ can be expressed in terms of the one with $\ell=0$,
\be\label{SL2-P1}
P_n^{(\ell)}(x_1,x_2) =
\vev{\Psi_n^{(0)}(z_1,z_2)|(-L_-)^\ell\e^{z_1{x_1}+z_2{x_2}}}= (x_1+x_2)^\ell\,
P_n^{(0)}(x_1,x_2) \, ,
\ee
which corresponds to the lowest weight and is given by
\be\label{SL2-Four}
P_n^{(0)}(x_1,x_2) = \vev{ (z_1-z_2)^n|\e^{z_1{x_1}+z_2{x_2}}} = \int
[\mathcal{D}z_1]_{j_1} \int [\mathcal{D}z_2]_{j_2}\, (\bar z_1 -\bar
z_2)^n\,\e^{z_1{x_1}+z_2{x_2}} \,.
\ee
Using the properties of the $SL(2)$ scalar product (see Appendix B), the
relations \re{SL2-P} -- \re{SL2-Four} can be inverted as~\cite{DerKehMan97}
\ba
\widebar{\Psi_n^{(\ell)}(z_1,z_2)} \!\!\!&=&\!\!\! P_n^{(\ell)}(\partial_{w_1},
\partial_{w_2})\prod_{k=1,2}\Gamma(2j_k)(1-w_k \bar z_k)^{-2j_k}\bigl|_{\bar w_k=0}
\nonumber \\
\label{PPsi}
&=&\!\!\! \int_0^\infty \prod_{k=1,2}{dt_k\, t_k^{2j_k-1}
} \e^{-t_1-t_2}P_n^{(\ell)}(t_1 \bar z_1,t_2 \bar z_2) \, .
\ea
Eqs.~\re{SL2-P} -- \re{PPsi} establish the correspondence between the conformal
polynomials defining the twist-two operators \re{OP} and the lowest weights in
the tensor product of two $SL(2)$ modules, Eqs.~\re{SL2-dec} and \re{SL2-0}.

According to \re{SL2-Four}, the polynomial $P_n^{(0)}(x_1,x_2)$ is given by the
lowest weight $\Psi_n^{(0)}(z_1,z_2)$ transformed from the ``coordinate''
$z-$representation to the ``momentum'' $x-$representation. Making use of
\re{SL2-norm}, the scalar product in \re{SL2-Four} can be evaluated and yields
\be
P_n^{(0)}(x_1,x_2) = \sum_{k=0}^n (-1)^{n-k} \left( n \atop k \right) \vev{ z_1^k
z_2^{n-k}|\e^{z_1{x_1}+z_2{x_2}}} = \sum_{k=0}^n  \frac{x_1^k (-x_2)^{n-k}\left(
n \atop k \right)}{\Gamma(2j_1+k)\Gamma(2j_2+n-k)}\,.
\ee
In turn, the sum can be expressed in terms on the Jacobi polynomials (see
Eq.~\re{Jacobi-def})
\ba
\label{Jacobi}
P^{(0)}_n(x_1,x_2) &=& a_n^{(2j_1,2j_2)}\cdot (x_1+x_2)^n \,{\rm
{P}}_n^{(2j_1-1,2j_2-1)} \left( \frac{x_2-x_1}{x_2+x_1} \right) \, ,
\ea
with $a_n^{(2j_1,2j_2)}=(-1)^n n!/[\Gamma(n+2j_1)\Gamma(n+2j_2)]$. Together with
\re{OP}, this relation leads to the well-known expression~\cite{Mak81} for the
twist-two conformal $SL(2)$ operator, $\mathcal{O}^{(0)}_n(0)\equiv {\mathcal
O}_n^{j_1,j_2} (0)$, built from fields of unequal conformal spins $j_1$ and $j_2$
\be\label{gen-def}
{\mathcal O}_n^{{j_1},{j_2}}(x) = \partial_+^n \tr \left[\phi_{j_1}(x)
\mathrm{P}^{(2j_1-1,2j_2-1)}_n \left( \frac{ \stackrel{\rightarrow}{\partial}_+ -
\stackrel{\leftarrow}{\partial}_+ }{ \stackrel{\rightarrow}{\partial}_+ +
\stackrel{\leftarrow}{\partial}_+} \right) \phi_{j_2}(x)\right] ,
\ee
which transforms under the $SL(2)$ transformations according to \re{SL2-trans}
with the conformal spin $j=j_1+j_2+n$.

The following comments are in order.

The operator \re{gen-def} was constructed from the lowest weight $P^{(0)}_N
(x_1,x_2)$. The twist-two operators corresponding to the polynomials
$P^{(\ell)}_N (x_1,x_2)$ with $\ell\ge 1$ differ from \re{gen-def} by total
derivatives $\mathcal{O}^{(\ell)}_n(0)= \partial_+^\ell {\mathcal
O}_n^{j_1,j_2}(0)$ and, therefore, have the same anomalous dimension.

For $j=j_1=j_2$, the Jacobi polynomials in \re{gen-def} are reduced to the
Gegenbauer polynomials, ${\mathrm{P}}_N^{(2j-1,2j-1)} (x) \sim
\mathrm{C}_N^{2j-1/2}(x)$ (see Eq.~\re{P=C}). In particular, in the
$\mathcal{N}=0$ theory, the twist-two operators are built from the gauge fields
$\partial_+ A$ and $\partial_+ \bar A$ of helicity $\pm 1$ which carry the same
conformal spin $j=3/2$. The corresponding Wilson operators are given by
${\mathcal O}_N^{{3/2},{3/2}}(0)$, Eq.~\re{gen-def}, and involve the same
conformal (Gegenbauer) polynomial $\mathrm{C}_N^{5/2}(x)$ (see Eq.~\re{GG}).
Still, the anomalous dimensions of these operators depend on the helicity
alignment of gauge fields, Eqs.~\re{H-PsiPsi-J} and \re{H-PhiPsi-J}.

So far we have implicitly assumed that the operators ${\mathcal O}_N^{j_1,j_2}(x)$
can not admix to other twist-two operators having a different ``partonic'' content.
Indeed, in the $\mathcal{N}=0$ theory (pure gluodynamics) the twist-two operators
are built from the gauge field strengths only. They have an autonomous scale
dependence simply because there are no other operators they could mix with. For
$\mathcal{N}\ge 1$ this property is lost due to a growing number of constituent
fields. In supersymmetric Yang-Mills theory the conformal operators \re{gen-def}
can be constructed from the gauge field, $\phi_{j=3/2} =(\partial_+A, \partial_+
\bar A)$, gaugino, $\phi_{j=1}= (\lambda^A, \bar \lambda_A)$, and scalars,
$\phi_{j=1/2}=\phi^{AB}$. As a result, these operators mix under renormalization
and the conformal symmetry alone does not allow one to resolve their mixing. Let
us address this case next.

\section{Superconformal operators}
\label{SuperSection}

In SYM theory with $\mathcal{N}$ supercharges, the full $SU(2,2|\mathcal{N})$
superconformal group reduces to its ``collinear'' $SL(2|\mathcal{N})$
subgroup when projected on the light-cone. This subgroup includes the $SL(2)$
transformations, Eq.~\re{SL2-trans}, as well as supersymmetric transformations
of fields and their superconformal generalizations (see Eqs.~\re{SL2-subgroup}
-- \re{V-plus}). Then, the Wilson operators in SYM theory can be classified
according to representations of the $SL(2|\mathcal{N})$ group. The operators
defined in this way have automous scale dependence since their mixing with
other operators is protected to one-loop order by the superconformal symmetry.

\subsection{Superconformal polynomials}

In the light-cone superspace formalism, the elementary component fields of SYM
theory arise as coefficients in the Taylor expansion of the light-cone superfields
$\Phi(Z)=\Phi(z n_\mu,\theta^A)$ and $\Psi(Z)=\Psi(z n_\mu,\theta^A)$ in powers of
Grassmann coordinates $\theta^A$ (with ${\scriptstyle A}=1,\ldots,\mathcal{N}$).
This suggests to generalize the nonlocal light-cone operators \re{phi-phi} in the
way of considering the products of two superfields $\tr \left[\Phi(Z_1) \Phi(Z_2)
\right]$, $\tr \left[\Psi(Z_1) \Psi(Z_2)\right]$ and $\tr \left[\Phi(Z_1) \Psi(Z_2)
\right]$. Their expansion in $\theta_{1,2}^A$ and $z_{1,2}$ generates all possible
twist-two operators in underlying SYM theory, Eq.~\re{O-def}, built from different
constituent fields
\be\label{Q-SYM}
\mathbb{O}(Z_1,Z_2) =\sum_{n,\ell} \Psi^{(\ell)}_n(Z_1,Z_2)\,{\cal
O}^{(\ell)}_n(0)\,,
\ee
where the coefficient functions $\Psi^{(\ell)}_n(Z_1,Z_2)$ are polynomials in
$Z_k=(z_k,\theta_k^A)$ with $k=1,2$. Similarly to Eq.\ \re{OP}, the local
operator ${\cal O}^{(\ell)}_n(0)$ can be extracted from the nonlocal operator
$\mathbb{O} (Z_1,Z_2)$ with a help of a projection polynomial $P_n (X_1,X_2)$,
Eq.~\re{SuperConfOper}, with $X_k = (x_k,\vartheta_{k, A})$. In comparison to the
$\mathcal{N}=0$ case, this polynomial depends on additional $2\mathcal{N}$ odd
variables $\vartheta_{1,2, A}$. The operator ${\cal O}^{(\ell)}_n(0)$ is given by
\re{SuperConfOper} with the corresponding polynomial carrying two indices $n$ and
$\ell$. As we will argue below, $n$ enumerates irreducible components in the
tensor product of two $SL(2|\mathcal{N})$ representations and $\ell$
parameterizes the states within a given $SL(2|\mathcal{N})$ module. For
$\mathcal{N}=0$ the relation \re{Q-SYM} coincides with \re{Q}.

As before, let us associate two sets of polynomials $\Psi^{(\ell)}_n(Z_1,Z_2)$
and $P^{(\ell)}_n(X_1,X_2)$ with a given twist-two operator ${\cal O}^{(\ell)}_n(0)$.
According to their definition, Eqs.~\re{SuperConfOper} and \re{Q-SYM}, they satisfy
the orthogonality condition analogous to Eq.\ \re{ort-1}
\be\label{ort-SYM}
P^{(\ell)}_j
\left( \partial_{Z_1}; \partial_{Z_2} \right)
\, \Psi^{(\ell')}_{j'}(Z_1,Z_2)\bigl|_{Z_1=Z_2=0}
=
\delta_{jj'} \delta_{\ell\ell'}
\, .
\ee
Remarkably enough, the nonlocal operator \re{Q-SYM} satisfies the renormalization
group equation which is superficially identical to \re{H}
\be
\label{H-SYM}
\mu\frac{d}{d\mu}\mathbb{O}(Z_1,Z_2) =- \frac{g^2 N_c}{8\pi^2}\left[ ( \mathbb{H}
+ 2 \gamma_\mathcal{N}^{\scriptscriptstyle (0)} ) \cdot
\mathbb{O}\right](Z_1,Z_2) + {\cal O}(g^4) \, .
\ee
The important difference with the $\mathcal{N}=0$ case is that the evolution
kernel $\mathbb{H}$ acts both on the bosonic $z_{1,2}$ and fermionic $\theta_{1,2}^A$
coordinates \cite{BelDerKorMan04}. Substituting \re{Q-SYM} into \re{H-SYM}, one
finds that the polynomials $\Psi^{(\ell)}_{n}(Z_1,Z_2)$ corresponding to the
operators ${\cal O}^{(\ell)}_n(0)$ have to satisfy the stationary Schr\"odinger
equation
\be\label{Sch-SYM}
\left[ \mathbb{H} \cdot \Psi^{(\ell)}_{n}\right](Z_1,Z_2)
=
E (n) \Psi^{(\ell)}_{n}(Z_1,Z_2) \, ,
\ee
with $\gamma^{\scriptscriptstyle (0)} (n) = E (n) + 2
\gamma_\mathcal{N}^{\scriptscriptstyle (0)}$. Notice that the functional
dependence of anomalous dimension $\gamma^{\scriptscriptstyle (0)} (n)$ on
$n$ differs for the $\Phi\Phi-$, $\Psi\Psi-$ and $\Phi\Psi-$sectors,
Eqs.~\re{H-PsiPsi-J} and \re{H-PhiPsi-J}, respectively.

Now we are in a position to demonstrate that the eigenfunctions
$\Psi^{(\ell)}_{n} (Z_1,Z_2)$ are uniquely determined by the superconformal
$SL(2|\mathcal{N})$ symmetry of the classical SYM Lagrangian in the light-cone
formalism.

\subsection{$SL(2|\mathcal{N})$ superconformal symmetry}

The light-cone superfields $\Phi(Z)$ and $\Psi(Z)$ define representations of the
$SL(2|\mathcal{N})$ group that we shall denote as $\mathbb{V}_{j_\Phi}$ and
$\mathbb{V}_{j_\Psi}$, respectively. In the light-cone formalism, the superfields
are transformed linearly under the superconformal $SL(2|\mathcal{N})$
transformations
\be\label{local}
\delta_G \Phi_j(Z) = G \, \Phi_j(Z)\,,
\ee
with the $SL(2|\mathcal{N})$ generators $G=\{L^\pm, L^0, {W}{}^{A,\pm}, {V}^\pm_{A},
B,{T}_B{}^A\}$ being the differential operators acting on the coordinates of superfields
as
\begin{equation}
\label{sl2}
\begin{array}{llll}
{L}^- = -\partial_z \, , \ & {L}^+ = 2 j\, z + z^2\partial_z + z \left( \theta\cdot
\partial_\theta \right) \, , \ & {L}^0 = j + z
\partial_z + \ft12\left( \theta\cdot \partial_\theta \right)
\, , \ &
\\ [3mm] {W}{}^{A,-} = \theta^A \, \partial_z \, , \ & {W}{}^{A,+} =
\theta^A [ 2j  +  z \partial_z +  \left( \theta\cdot
\partial_\theta \right) ] \, , \ & {V}^-_{A} = \partial_{\theta^A} \, , \ &
{V}^+_{A} = z\partial_{\theta^A} \, , \!\!\!\!\\ [3mm] \, \ & {T}_B{}^A = \theta^A
\partial_{\theta^B} - \ft1{\mathcal{N}} \, \delta_B^A \left( \theta\cdot
\partial_\theta \right) \, , \ & {B} = - j - \ft12 \left( 1 - \ft{2}{{\cal N}}
\right) \left( \theta\cdot \partial_\theta \right) \, , \ &
\end{array}
\end{equation}
with $\partial_z \equiv \partial/\partial z$ and $\theta \cdot \partial_\theta \equiv
\theta^A \partial/\partial\theta^A$. Here the parameter $j$ is the superconformal spin
of the superfield. In SYM theory with $\mathcal{N}$ supercharges, the $\Phi-$ and
$\Psi-$superfields, Eqs.~\re{M=0-field} -- \re{M=4-field}, correspond to $j=-1/2$ and
$j=(3-\mathcal{N})/2$, respectively. A global form of the transformations \re{local}
can be found in Appendix~B, Eqs.~\re{SL2-subgroup} -- \re{V-plus}.

The representation $\mathbb{V}_{j}$ is spanned by polynomials in both $z$ and
$\theta^A$ which arise from the Taylor expansion of the superfields $\Phi_j(z
n_\mu,\theta^A)$ around $z=\theta^A=0$. It possesses the lowest weight $1$ which
is annihilated by the lowering operators ${L}^-$, ${W}{}^{A,-}$, ${V}^-_{A}$ and
satisfies the chirality condition $(L^0+B)\cdot 1 = 0$. For $j\ge 1/2$ the
representation $\mathbb{V}_{j}$ is irreducible and it is known in the literature
as atypical, or chiral representation \cite{FraSorSci,book}.

The product of two light-cone superfields $\mathbb{O}(Z_1,Z_2)$ belongs to the
tensor product of two chiral representations,
{$\mathbb{V}_{j_1}\otimes\mathbb{V}_{j_2}$}. The $SL(2|\mathcal{N})$ generators
on this tensor product are given by the sum of differential operators acting on
coordinates of two superfields, $G=G_{(1)}+G_{(2)}$. The superconformal
invariance implies that the one-loop evolution kernel $\mathbb{H}$ entering
\re{H-SYM} commutes with the $SL(2|\mathcal{N})$ generators, $[\mathbb{H},G]=0$.
Therefore, $\mathbb{H}$ depends on the two-particle superconformal spin
$\mathbb{J}_{12}$ defined as
\be\label{J_ab}
\mathbb{L}_{12}^2=\mathbb{J}_{12}(\mathbb{J}_{12}-1)+C_{12}\,,
\ee
with $C_{12}=\mathcal{N}(j_1+j_2)[1+(j_1+j_2)/(\mathcal{N}-2)]$ and
$\mathbb{L}_{12}^2$ being the two-particle quadratic Casimir operator of
the $SL(2|\mathcal{N})$ group
\be\label{Casimir}
\mathbb{L}_{12}^2
=
(L^0)^2 + L^+L^- + (\mathcal{N} - 1)L^0
+ \frac{\mathcal{N}}{\mathcal{N} - 2} B^2 -{V}^+_{A}{W}{}^{A,-}
- {W}{}^{A,+}{V}^-_{A} - \frac12{T}_B{}^A{T}_A{}^B
\, .
\ee
To construct the eigenfunctions of the $SL(2|\mathcal{N})$ invariant operator
$\mathbb{H}$, Eq.~\re{Sch-SYM}, it suffices to decompose the tensor product
{$\mathbb{V}_{j_1}\otimes\mathbb{V}_{j_2}$} over the irreducible
$SL(2|\mathcal{N})$ components. The decomposition takes the form
\cite{FraSorSci,book,Der00}
\be\label{SL2N-dec}
\mathbb{V}_{j_1}\otimes\mathbb{V}_{j_2} = \sum_{n\ge 0}
\mathcal{V}_{j_1+j_2+n}\,,
\ee
where the sum in the right-hand side goes over the $SL(2|\mathcal{N})$
representations with the superconformal spin $\mathbb{J}_{12}=j_1+j_2+n$.

The eigenfunctions $\Psi^{(\ell)}_{n}(Z_1,Z_2)$, Eq.~\re{Sch-SYM}, belong to the
$SL(2|\mathcal{N})$ module $\mathcal{V}_{j_1+j_2+j}$ and can be classified as
follows. For $\ell=0$, the polynomial $\Psi_{n}^{(0)}(Z_1,Z_2)$ is the lowest
weight of $\mathcal{V}_{j_1+j_2+n}$. By definition, it is annihilated by the
lowering $SL(2|\mathcal{N})$ generators
\be L^-\, \Psi_n^{(0)}(Z_1,Z_2) = W^{A,-}
\,\Psi_n^{(0)}(Z_1,Z_2) = V_A^-\, \Psi_n^{(0)}(Z_1,Z_2) = 0
\ee
and carries the superconformal spin \re{J_ab} equal to $j_1+j_2+n$,
\be
\mathbb{J}_{12}\, \Psi_n^{(0)}(Z_1,Z_2) =  (j_1+j_2+n)\Psi_n^{(0)}(Z_1,Z_2)\,.
\ee
Depending on the value of the nonnegative integer $n$, the lowest weight is given
by the following expressions
\be\label{weights}
\Psi_{n}^{(0)}(Z_1,Z_2)
=
\left\{
\begin{array}{ll}
1 \, , & n=0 \\[2mm]
\varepsilon_{A_1\ldots A_n A_{n+1}\ldots A_\mathcal{N}}
\theta_{12}^{A_1}\ldots \theta_{12}^{A_n}
\, , & 1< n < \mathcal{N} \\[2mm]
\varepsilon_{A_1\ldots A_\mathcal{N}}
\theta_{12}^{A_1} \ldots \theta_{12}^{A_\mathcal{N}}\cdot (z_1-z_{2})^{n-\mathcal{N}}
\, , & n \ge \mathcal{N} \\
\end{array}
\right.
\ee
where $\theta_{12}^{A}=\theta_{1}^{A}-\theta_{2}^{A}$. For $\ell \ge 1$, the
polynomials $\Psi_n^{(\ell)}(Z_1,Z_2)$ are obtained from the lowest weight
\re{weights} by applying the raising $SL(2|\mathcal{N})$ generators
$L^+,{W}{}^{A,+}, {V}^+_{A}$ and ${T}_B{}^A$. To save space we do not present
their explicit form here.

The lowest weight $\Psi_n^{(0)}(Z_1,Z_2)$, Eq.~\re{weights}, uniquely specifies
the representation $\mathcal{V}_{j_1+j_2+n}$ in the right-hand side of
\re{SL2N-dec}. One can verify that the polynomial $\Psi_n^{(0)}(Z_1,Z_2)$
diagonalizes the generators $B$ and $L^0$
\ba
B\,\, \Psi_n^{(0)}(Z_1,Z_2) \!\!\!&=&\!\!\! -
\left(j_1+j_2+\frac{\mathcal{N}-2}{2\mathcal{N}}n_<\right) \Psi_n^{(0)}(Z_1,Z_2)
\, ,
\\
L^0\, \Psi_n^{(0)}(Z_1,Z_2) \!\!\!&=&\!\!\! \left(j_1+j_2+n-\frac12n_<\right)
\Psi_n^{(0)}(Z_1,Z_2)\,,
\ea
with $n_< = {\rm min} (n,\mathcal{N})$ and  carries (for $\mathcal{N} \ge 2$) a
nontrivial $SU(\mathcal{N})$ charge
\be
\left({T}_B{}^A{T}_A{}^B\right)\, \Psi_n^{(0)}(Z_1,Z_2) =t_n\,
\Psi_n^{(0)}(Z_1,Z_2)\,,
\ee
with $t_n={n(\mathcal{N}-n)(\mathcal{N}+1)}/{\mathcal{N}}$ for $0\le n\le
\mathcal{N}$ and $t_n=0$ for $n > \mathcal{N}$. For $n = 0$ the lowest weight
$\Psi_{n=0}^{(0)}(Z_1,Z_2)=1$ satisfies the chirality condition
\be
(B+L^0)\Psi_{0}^{(0)}(Z_1,Z_2)=0
\ee
and, as a consequence the corresponding $SL(2|\mathcal{N})$ representation
$\mathcal{V}_{j_1+j_2}$, Eq.~\re{SL2N-dec} is atypical, or chiral. For $n\ge 1$
the representation $\mathcal{V}_{j_1+j_2+n}$ is typical.

The polynomials $\Psi_n^{(\ell)}(Z_1,Z_2)$ defined in this way diagonalize the
one-loop Hamiltonian $\mathbb{H}$ \re{Sch-SYM}. They determine the coefficient
functions accompanying the twist-two operators in the OPE expansion \re{Q}. By
virtue of the $SL(2|\mathcal{N})$ invariance, the anomalous dimension of these
operators is a function of the superconformal spin $j_1+j_2+n$ and does not
depend on $\ell$.

\subsection{Invariant scalar product}

Let us establish a relation between the eigenstates $\Psi_n^{(\ell)}(Z_1,Z_2)$,
determined in the previous section, and the polynomials $P_n(X_1;X_2)$ in $X =
(x, \vartheta_A)$ defining the superconformal operators, Eq.~\re{SuperConfOper}.
The consideration goes along the same lines as in Sect.~2.2 for $\mathcal{N}=0$.
Namely, we shall introduce an $SL(2|\mathcal{N})$ invariant scalar product on the
space spanned by the polynomials in $Z_1$ and $Z_2$ and, then, project both sides
of \re{Q-SYM} onto the lowest weight $\Psi_n^{(0)}(Z_1,Z_2)$.

To begin with let us consider the $\mathcal{N}=1$ case. An arbitrary polynomial
$\Psi_j(Z)$ belonging to the chiral $SL(2|1)$ module $\mathbb{V}_j$ can be
expanded in powers of $\theta$ as
\be\label{vector}
\Psi_j(z,\theta)=\psi(z)+\theta\,\chi(z)\,,
\ee
with $\psi(z)$ and $\chi(z)$ being some polynomials in $z$. Since the $SL(2|1)$
group contains the $SL(2)$ subgroup generated by the bosonic operators $L^\pm$
and $L^0$, Eq.~\re{sl2}, the polynomials $\psi(z)$ and $\chi(z)$ are transformed
under the $SL(2)$ transformations according to Eq.\ \re{SL2-trans} and carry a
definite value of the $SL(2)$ spins, $j$ and $j+1/2$, respectively. We remind
that for an arbitrary (half-)integer spin $j\ge 1/2$, the $SL(2)$ scalar product
on the space spanned by polynomials in $z$ is given by \re{sc}. This suggests to
define the scalar product for the states \re{vector} as
\be\label{c-const}
\vev{\Psi|\Psi'}_{\scriptscriptstyle SL(2|1)} = \vev{\psi|\psi'}_j + c\,
\vev{\chi|\chi'}_{j+\frac12}\,,
\ee
where $c$ is an arbitrary constant and the subscript in the right-hand side
indicates the value of the $SL(2)$ conformal spin in Eq.~\re{sc}.

By construction, Eq.\ \re{c-const} is invariant under the $SL(2)$ transformations
\re{SL2-trans} for abritrary $c$. The value of $c$ has to be fixed by imposing
the condition of invariance of \re{c-const} under the $SU(1,1|1)$ transformations,
Eqs.~\re{SL2-subgroup} -- \re{V-plus}. To this end, it proves convenient to
rewrite \re{c-const} in terms of the $\mathcal{N}=1$ superfields \re{vector}.
It is straightforward to verify that for the states $\Psi(Z)$ and $\Psi'(Z)$
belonging to the $SL(2|1)$ module $\mathbb{V}_j$ the following integral is
invariant under the superconformal transformations (Eqs.~\re{SL2-subgroup} --
\re{V-plus})
\be\label{ansatz}
\vev{\Psi|\Psi'}_{\scriptscriptstyle SL(2|1)}=\frac1{\pi\Gamma(2j)} \int_{|z|\le
1} d^2 z\int d\bar\theta d\theta\, (1-z\bar z-\theta
\bar\theta)^{2j-1}\widebar{\Psi(Z)} \,\Psi'(Z)\,.
\ee
Here the integration over odd coordinates is performed using the identities
\be
\label{theta-rules}
\int d\theta =\int d\bar\theta =0\,,\qquad \int d\theta\,\theta =-\int d\bar
\theta\,\bar\theta =1\,,
\ee
which is consistent with the complex conjugation assumed for odd variables
$\widebar{(\chi\theta)} = \bar\theta\bar\chi$ and $\bar Z=(\bar z,\bar\theta)$.
Expanding the integrand in Eq.\ \re{ansatz} in powers of $\theta$ and
$\bar\theta$ one arrives at \re{c-const} with $c=1$. Eq.~\re{ansatz} defines
the $SL(2|1)$ scalar product on the space $\mathbb{V}_j$ spanned by the
$\mathcal{N}=1$ superfields \re{vector}.

The above consideration can be easily generalized to $\mathcal{N}\ge 2$. In that
case, the expansion in the right-hand side of \re{vector} goes in powers of
$\theta^{A}$ and involves $2^{\mathcal{N}}$ different polynomials which carry
both the $SU(\mathcal{N})$ isotopic charge and the $SL(2)$ conformal spin.
Similar to \re{c-const}, the $SL(2|\mathcal{N})$ scalar product is given by the
sum over the $SL(2)$ scalar products of the conformal spins varying between $j$
and $j+\mathcal{N}/2$. It takes a particularly simple form when written as an
integral over the superspace
\be\label{ansatz-N}
\vev{\Psi|\Psi'}_{\scriptscriptstyle SL(2|\mathcal{N})} = \int [\mathcal{D}Z
]_{j }\, \widebar{\Psi(Z)} \Psi'(Z) \,,
\ee
where the notation was introduced for the integration measure
\be
\label{measure-N}
\int [\mathcal{D}Z ]_{j}=\frac1{\pi\Gamma(2j-1+\mathcal{N})}\int_{|z|\le 1} d^2
z\int \prod_{A=1}^\mathcal{N} \left(d\bar\theta_A d\theta^A \right)\, (1-z\bar
z-\theta\cdot
\bar\theta)^{2j+\mathcal{N}-2}\,,
\ee
with $Z=(z,\theta^A)$, $\bar Z=(\bar z,\bar\theta_A)$ and $\theta\cdot\bar\theta
\equiv \sum_{A=1}^\mathcal{N} \theta^A \bar\theta_A$. For $\mathcal{N}=0$ and
$\mathcal{N}=1$ these relations match Eqs.~\re{sc} and \re{ansatz}, respectively.
For $\mathcal{N} \geq 2$, the integration measure over Grassmann variables is
normalized via
\be
\int \prod_{A=1}^\mathcal{N}  d\theta^A \cdot \theta^{A_1} \cdots
\theta^{A_{\mathcal N}} = \varepsilon^{A_1 \dots A_{\mathcal{N}}} \, ,
\ee
with $\varepsilon^{12\dots\mathcal{N}} = 1$. One can verify that the scalar
product \re{ansatz-N} is invariant under the superconformal $SU(1,1|\mathcal{N})$
transformations, Eqs.~\re{SL2-subgroup} -- \re{V-plus}.

The $SL(2|\mathcal{N})$ representation space $\mathbb{V}_j$ is spanned by the
states $\Psi(Z) \sim z^n \theta^{A_1} \ldots \theta^{A_L}$ given by polynomials
in both even and odd variables. They are orthogonal with respect to the scalar
product \re{ansatz-N}, namely,
\be\label{norm-N}
\vev{z^n \theta^{A_1}\ldots\theta^{A_L}| z^k\theta^{B_1}\ldots\theta^{B_M}}
=\delta_{nk} \delta_{LM} \frac{n!}{\Gamma(2j+n+L)}
\left(\delta_{A_1}^{B_1}\ldots\delta_{A_L}^{B_L} + \cdots \right) \,,
\ee
with ellipses standing for the terms which ensure antisymmetry of the right-hand
side with respect to the permutation of any pair of lower/upper indices. For the
polynomials $\Psi(Z_1,Z_2)$ belonging to the tensor product $\mathbb{V}_j\otimes
\mathbb{V}_j$, the scalar product is given by the integral over $Z_1-$ and
$Z_2-$coordinates in the superspace with the same measure as in \re{ansatz-N}.

The $SL(2|\mathcal{N})$ generators, $L^0, B$ and ${T^A}_B$, Eqs.~\re{sl2},  are
self-adjoint operators with respect to the scalar product \re{ansatz-N}, that is
$\vev{\Psi L^0|\Psi'}=\vev{\Psi|L^0\Psi'}$. For the remaining $SL(2|\mathcal{N})$
generators one finds
\be\label{conj-N}
(L^+)^\dagger=-L^-\,,\qquad (W^{A,\pm})^\dagger=V_A^\mp\,,
\ee
and, as a consequence, the Casimir operator \re{Casimir} is a self-adjoint
operator on the tensor product \re{SL2N-dec},
$(\mathbb{L}_{12}^2)^\dagger=\mathbb{L}_{12}^2$. Therefore, its eigenfunctions
$\Psi_{n}^{(\ell)}(Z_1,Z_2)$, Eq.~\re{weights}, are mutually orthogonal with
respect to \re{ansatz-N},
\be\label{norm1-N}
\vev{\Psi_{n}^{(\ell)}(Z_1,Z_2) |\Psi_{k}^{(\ell')}(Z_1,Z_2) } \sim \delta_{nk}
\delta_{\ell \ell'}\,.
\ee
This relation allows one to determine the explicit form of superconformal
operators in SYM theories.

Let us project both sides of Eq.\ \re{Q-SYM} onto $\Psi_{n}^{(\ell)}(Z_1,Z_2)$.
Taking into account \re{norm1-N}, one obtains the following representation for
the twist-two operators in the $\mathcal{N}-$extended SYM
\be\label{N-v}
\mathcal{O}_n^{(\ell)}(0) \sim \vev{\Psi_n^{(\ell)}(Z_1,Z_2)|
\mathbb{O}(Z_1,Z_2)} =\int [\mathcal{D}Z_1]_{j_1}\int
[\mathcal{D}Z_1]_{j_2}\,\overline{\Psi^{(\ell)}_{n}(Z_1,Z_2)}\,
\,\mathbb{O}(Z_1,Z_2) \, .
\ee
This relation is a natural generalization of a similar relation for $\mathcal{N}=0$,
i.e., Eq.~\re{N=0-v}. Expanding $\mathbb{O}(Z_1,Z_2)$ in the Taylor series around
$Z_1=Z_2=0$ and matching \re{N-v} into \re{SuperConfOper}, we identify the
corresponding superpolynomials $P_n^{(\ell)}$ as
\ba\label{SL2N-F}
P_n^{(\ell)}(X_1,X_2) &=& \vev{\Psi_n^{(\ell)}(Z_1,Z_2)| \textrm{e}^{\,Z_1\cdot
X_1+Z_2\cdot X_2}} \\
\nonumber &=& \int [\mathcal{D}Z_1]_{j_1}\int
[\mathcal{D}Z_1]_{j_2}\,\overline{\Psi^{(\ell)}_{n}(Z_1,Z_2)}\,
\,\textrm{e}^{\,Z_1\cdot X_1+Z_2\cdot X_2}\,.
\ea
Here the notation was introduced for the momentum variables in the superspace
$X_k=(x_k,\vartheta_{k,A})$
\be
Z_k\cdot X_k \equiv z_k x_k +\sum_{A=1}^\mathcal{N} \theta_{k}^A\,\vartheta_{k,A}
\,.
\ee
In full analogy with Eq.~(\ref{PPsi}), the inverse transformation looks like
(see Eqs.~\re{id} and \re{id1})
\ba
\widebar{\Psi_n^{(\ell)}(Z_1,Z_2)}
&=&  P_n^{(\ell)}(\partial_{w_1},\partial_{\vartheta_{1}^A};
\partial_{w_2},\partial_{\vartheta_{2}^A})
\prod_{k=1,2}\Gamma(2j_k)(1-w_k \bar z_k -
\vartheta\cdot\bar\theta)^{-2j_k}\bigl|_{ w_k=\vartheta^A_k=0}
\nonumber\\
\label{PPsi-S}
&=&  \int_0^\infty \prod_{k=1,2} {dt_k\, t_k^{2j_k-1} }
\e^{-t_1-t_2}
P_n^{(\ell)}
(t_1 \bar z_1, t_1\bar \theta_{1,A_1};t_2 \bar z_2,t_2 \bar\theta_{2,A_2})
\,.
\ea
It is instructive to compare this relation with \re{PPsi}. One notices that
the only difference is that going over from $\mathcal{N}=0$ to $\mathcal{N}\ge 1$
one has to enlarge the number of odd directions in the superspace.

Similar to \re{SL2-P}, the polynomials $P_n^{(\ell)}(X_1,X_2)$ for $\ell \geq 1$
can be expressed in terms of the one with $\ell=0$. The descendant eigenstates
$\Psi_n^{(\ell)}(Z_1,Z_2)$ are obtained from the lowest weight $\Psi_n^{(0)}
(Z_1,Z_2)$ by applying the raising operators $L^+,{W}{}^{A,+},{V}^+_{A}$ and
${T}_B{}^A$. Taking into account \re{SL2N-F} and \re{conj-N}, these operators can
be realized as differential operators $\widehat{L}^+,\widehat{W}{}^{A,+},
\widehat{V}^+_{A}$ and $\widehat{T}_B{}^A$ acting on the polynomial $P_n^{(0)}
(X_1,X_2)$. For example, the operator $\widehat{W}{}^{A,+}$ is defined as
\ba
 \lefteqn{\vev{{V}^+_{A}\Psi_n^{(0)}(Z_1,Z_2)| \textrm{e}^{\,Z_1\cdot X_1+Z_2\cdot
X_2}} = \vev{\Psi_n^{(0)}(Z_1,Z_2)|{W}{}^{A,-} \textrm{e}^{\,Z_1\cdot
X_1+Z_2\cdot X_2}}} \\[2mm]
&= &(-1)^{\bar n} \sum_{k=1,2} \left(-x_k\partial_{\vartheta_{k,A}}
 \right)\vev{\Psi_n^{(0)}(Z_1,Z_2)| \textrm{e}^{\,Z_1\cdot X_1+Z_2\cdot
X_2}}\equiv  (-1)^{\bar n}\widehat{V}^+_{A} P_n^{(0)}(X_1,X_2)\,, \nonumber
\ea
with the grading factor $(-1)^{\bar n}$ equal to $1$ (or $-1$) for polynomials
involving even (or odd) number of grassman variables. In this way, one finds\\[0mm]
\ba\label{hat-ops}
&& \widehat{V}^+_{A}  = \sum_{k=1,2}\left(-x_k\partial_{\vartheta_{k,A}}
\right)\,,\qquad \widehat{W}^{A,+} = \vartheta_{1,A} + \vartheta_{2,A}\,,\qquad
\\
&& \widehat{L}^+ = x_1 + x_2\,, \qquad\qquad\qquad  \widehat{T}_B{}^A =
\sum_{k=1,2}\vartheta_{k,B}
\partial_{\vartheta_{k,A}} - \frac1{\scriptstyle \mathcal{N}} \delta_B^A (\vartheta_k\cdot
\partial_{\vartheta_k})\,.
\nonumber
\ea\\[-2mm]
To evaluate the polynomials $P_n^{(\ell)}(X_1,X_2)$, it suffices to apply the
raising operators $\widehat{L}^+,\widehat{W}{}^{A,+},\widehat{V}^+_{A}$ and
$\widehat{T}_B{}^A$ to the lowest weight $P_n^{(0)}(X_1,X_2)$. This allows us to
restrict the consideration to the lowest weights $P_n^{(0)}(X_1,X_2)$ only.

Eqs.~\re{SL2N-F} and \re{PPsi-S} represent the main result of the paper. They
relate to each other the $SL(2|\mathcal{N})$ lowest weights, Eq.~\re{weights},
and the polynomials defining the superconformal twist-two operators in the SYM
theory with an arbitrary number of supercharges $\mathcal{N}$. As we will show in
Sect.~4, the resulting Wilson operators coincide with known expressions obtained
through diagonalization of the one-loop mixing matrix for $\mathcal{N}=1,2,4$.

\section{Twist-two operators in SYM}

Let us apply \re{SL2N-F} and \re{PPsi-S} to reconstruct the superconformal
operators out of the lowest weights \re{weights} in the $\Psi\Psi-$, $\Phi\Psi-$
and $\Phi\Phi-$sectors. We remind that for $\mathcal{N}=0$ a similar integral
transformation, Eq.~\re{PPsi}, relates the Jacobi polynomials, Eq.~\re{Jacobi},
to the $SL(2)$ lowest weights, Eq.~\re{SL2-0}.

The lowest weights \re{weights} are factorized into the product of two
translation invariant polynomials depending separately on even and odd
coordinates. Substituting \re{weights} into \re{SL2N-F} one can perform $Z_1-$
and $Z_2-$integrations by expanding the $SL(2|\mathcal{N})$ integration measure
\re{ansatz-N} in powers of $(\theta\cdot\bar\theta)$
\be\label{dec1}
\int [\mathcal{D} Z]_j = \sum_{k=0}^\mathcal{N} \int [\mathcal{D} z]_{j+k/2} \int
\prod_{A=1}^\mathcal{N} \left(d\bar\theta_Ad\theta^A \right)\,
\frac{(\bar\theta\cdot \theta)^{\mathcal{N}-k}}{(\mathcal{N}-k)!}\,,
\ee
with the $SL(2)$ measure $[\mathcal{D} z]_{j+k/2}$ defined in \re{SL2-measure}.
Then, for a test function of the same form as the lowest weight \re{weights},
\be
\label{Psi-ansatz}
\Psi(Z_1,Z_2) = (z_1-z_2)^N \varphi_M(\theta_1-\theta_2)\,,
\ee
with $\varphi_M(\theta)$ being a homogenous polynomial in $\theta^A$ of degree
$M$ (such that $M \le \mathcal{N}$), the integral in the right-hand side of
\re{SL2N-F} can be factorized into the product of $z-$ and $\theta-$integrals.
The $z-$integral is the same as for $\mathcal{N}=0$, Eq.~\re{SL2-Four}, and is
given by the Jacobi polynomial \re{Jacobi}. The $\theta-$integral can be easily
evaluated with a help of \re{theta-rules} leading to (see Eq.~\re{new})
\ba
\label{P-ansatz}
P(X_1,X_2) &=& (x_1+x_2)^N\sum_{{k_1,k_2\ge 0}\atop k_1+k_2=M}
{c_N^{(2j_1+k_1,2j_2+k_2)}}\cdot  {\rm {P}}_N^{(2j_1+{k_1}-1,2j_2+{k_2}-1)}
\left( \frac{x_2-x_1}{x_2+x_1} \right) \nonumber
\\
&& \qqquad \times \left(\vartheta_1\cdot{\partial_{\bar\theta_1}}\right)^{k_1}
\left(\vartheta_2\cdot{\partial_{\bar\theta_2}} \right)^{k_2}
\widebar{\varphi_M(\theta_1-\theta_2)} \bigg|_{\bar\theta_{1,2}=0}\,,
\ea
with the expansion coefficients
\be\label{cc}
c_N^{(2j_1+k_1,2j_2+k_2)}= (-1)^N
N!/[\Gamma(N+2j_1+k_1)\Gamma(N+2j_2+k_2)\,k_1!\, k_2!]\,.
\ee
It follows from \re{P-ansatz} that the superconformal polynomials corresponding
to the lowest weights \re{weights} are given by the sum of the $SL(2)$ conformal
polynomials multiplied by the product of $\vartheta-$variables.

In Sect.~3, it was tacitly assumed that the superfield $\Psi_j(Z)$ belongs to
an {\it irreducible} representation of the superconformal group. This implies
that its superconformal spin has to satisfy the condition $j\ge 1/2$. We remind
that in the light-cone formalism the superconformal spins of the superfields
$\Phi(Z)$ and $\Psi(Z)$, Eqs.~\re{M=0-field} -- \re{M=4-field}, are
$j_\Phi=-1/2$ and $j_\Psi=(3-\mathcal{N})/2$, respectively. For $\mathcal{N}<4$
the condition $j\ge 1/2$ is satisfied only for the $\Psi-$superfield and,
therefore, Eqs.~\re{SL2N-F} and \re{PPsi-S} can be safely applied in the
$\Psi\Psi-$sector. In order to construct superconformal polynomials in
$\Psi\Phi-$ and $\Phi\Phi-$sectors, we have to extend our consideration to
{\it reducible} representations of the $SL(2|\mathcal{N})$ group. This will be
done in Sect.~4.2.

\subsection{Superconformal operators in the $\Psi\Psi-$sector}

In SYM theories with $\mathcal{N}=1$ and $\mathcal{N}=2$ supercharges, the
twist-two operators in this sector are defined as
\be
\label{O-N=1}
\mathcal{O}_n^{(0)}(0)
=
P_n^{(0)}(\partial_{Z_1}; \partial_{Z_2})
\tr \left[\Psi(Z_1) \Psi(Z_2)\right]\bigg|_{Z_1 = Z_2 = 0}
\, ,
\ee
in terms of the light-cone superfield $\Psi(Z)$ given by Eqs.~\re{M=1-field} and
\re{M=2-field}, respectively. By construction, the operators \re{O-N=1} are the
lowest weights in the tensor product of two $SL(2|\mathcal{N})$ representations
$\mathbb{V}_{j_\Psi}\otimes \mathbb{V}_{j_\Psi}$.

\subsubsection*{$\mathcal{N}=1$ theory}

The superfield $\Psi(Z)$, Eq.~\re{M=1-field}, describes the negative helicity
components of the gauge $(\partial_+ \bar A)$ and gaugino $(\lambda)$ fields. It
carries the superconformal spin $j_\Psi=1$ and defines the irreducible
representation of the $SL(2|1)$ group, $\Psi(Z)\in \mathbb{V}_1$. The nonlocal
light-cone operator $\mathbb{O}(Z_1,Z_2) = \tr \left[\Psi(Z_1)\Psi(Z_2) \right]$
belongs to the tensor product $\mathbb{V}_1\otimes \mathbb{V}_1$ and its
expansion around $Z_1=Z_2=0$ produces an infinite set of twist-two operators.
According to \re{weights}, these operators are in the one-to-one correspondence
with the $SL(2|1)$ lowest weights
\ba
\label{Psi-N1}
\Psi_0^{(0)}(Z_1,Z_2)&=&1\,,
\\[2mm]
\Psi_n^{(0)}(Z_1,Z_2)&=&(\theta_1-\theta_2) (z_1-z_2)^{n-1}\,, \nonumber
\ea
where $n\ge 1$ and $Z=(z,\theta)$. Matching these expressions into
\re{Psi-ansatz} and \re{P-ansatz} one finds that the corresponding superconformal
polynomials are given by
\ba
\label{SCO-N1}
P_0^{(0)}( {X}_1; {X}_2) &=& 1\,,
\\
P_n^{(0)}( {X}_1; {X}_2) &=& c_n
\cdot
(x_1+x_2)^{n-1}
\left\{\vartheta_1\,\mathrm{P}_{n-1}^{(2,1)}\left(\frac{x_2-x_1}{x_2+x_1}\right)
-\vartheta_2\,\mathrm{P}_{n-1}^{(1,2)}\left(\frac{x_2-x_1}{x_2+x_1}\right)\right\}
\,,
\nonumber
\ea
where ${X}_k=(x_k,\vartheta_k)$ and $c_n=(-1)^{n-1}/[n\,(1+n)!]$. It is
straightforward to verify that these polynomials are related to the lowest
weights \re{Psi-N1} through the integral transformation \re{PPsi-S}.

Let us substitute \re{SCO-N1} into \re{O-N=1} and take into account that ${\Psi}
(z, \theta) = - \lambda(x) + \theta\partial_+
\bar A(x)$, Eq.~\re{M=1-field}:
\begin{itemize}
\item  For $n=0$ the twist-two operator vanishes
\be
\label{O1-N=1}
\mathcal{O}_0^{(0)}(0) = \tr \left[\Psi(0)\Psi(0) \right] =\tr
\left[\lambda(0)\lambda(0) \right] = 0\,.
\ee
\item  For $n\ge 1$, one finds (up to an overall normalization factor)
\ba
\nonumber \mathcal{O}_n^{(0)}(0) &=&  \tr \left[\partial_+ \bar A(0)\,
\mathbb{P}^{(2,1)}_{n-1}(\stackrel{\rightarrow}{\partial}_+ ,
\stackrel{\leftarrow}{\partial}_+) \,\lambda(0) + \lambda(0)\,
\mathbb{P}^{(1,2)}_{n-1}(\stackrel{\rightarrow}{\partial}_+ ,
\stackrel{\leftarrow}{\partial}_+)\,
\partial_+ \bar A(0) \right]
\\
&=& 2 \sigma_{n-1} \tr \left[\partial_+ \bar A(0)\,
\mathbb{P}^{(2,1)}_{n-1}(\stackrel{\rightarrow}{\partial}_+ ,
\stackrel{\leftarrow}{\partial}_+) \,\lambda(0)\right]\,,
\label{O2-N=1}
\ea
\end{itemize}
with $\sigma_{n-1}=[1+(-1)^{n-1}]/2$. Here the notation was introduced for the
differential operator
\be\label{P-op}
\mathbb{P}^{(a,b)}_{n}(\stackrel{\rightarrow}{\partial}_+ ,
\stackrel{\leftarrow}{\partial}_+)  = \mathrm{P}^{(a,b)}_{n} \left( \frac{
\stackrel{\rightarrow}{\partial}_+ - \stackrel{\leftarrow}{\partial}_+ }{
\stackrel{\rightarrow}{\partial}_+ + \stackrel{\leftarrow}{\partial}_+}
\right)( \stackrel{\rightarrow}{\partial}_+ +
\stackrel{\leftarrow}{\partial}_+)^{n}\,.
\ee
The operators $\mathcal{O}_n^{(0)}(0)$ carry the superconformal spin $j = 2j_\Psi
+ n=2+n$ and have autonomous scale dependence to one-loop order. Their anomalous
dimension is given by $E_{\Psi\Psi}(j) = 2\left[ \psi(n+2)-\psi(1)\right]$,
Eqs.~\re{gg} and \re{H-PsiPsi-J}. It is instructive to examine \re{O2-N=1} in
the forward limit (fw), that is, neglecting operators containing total derivatives.
Setting $\stackrel{\rightarrow}{\partial}_+ + \stackrel{\leftarrow}{\partial}_+=0$,
one finds from \re{O2-N=1} with a help of \re{as}
\be
\mathcal{O}_n^{(0)}(0)\,\, \stackrel{\rm fw}{\sim } \,\,\sigma_{n-1} \tr
\left[\partial_+
\bar A(0)\, {\partial}_+^{n-1} \,\lambda(0)\right],
\ee
which is only different from zero for odd $n$. This relation gives one of the
operators found in Ref.~\cite{BelDerKorMan04} by direct diagonalization of the
mixing matrix.

The operator \re{O2-N=1} belongs to the $\mathcal{N}=1$ supermultiplet of
twist-two operators with the aligned helicity. The remaining components of
this supermultiplet are determined by descendants of the lowest weights
\re{Psi-N1}. They are obtained by acting with the (two-particle) raising
operators $L^+$, $W^+$ and $V^+$, Eq.~\re{sl2}, on the lowest weight
$\Psi^{\scriptstyle (0)}_n (Z_1, Z_2)$. For example, $L^+\Psi^{\scriptstyle
(0)}_n (Z_1, Z_2)$ yields a linear combination of $(z_1 + z_2) (\theta_1 -
\theta_2) (z_1 - z_2)^{n - 1}$ and $(\theta_1 + \theta_2) (z_1 - z_2)^n$. When
transformed into superconformal operators, the former gives a total derivative
of the operator \re{O2-N=1}, while the latter produces the other parity
component of the supermultiplet in question
$$
\sigma_{n-1} \tr \left[\partial_+ \bar A(0)\,
\mathbb{P}^{(2,1)}_{n}(\stackrel{\rightarrow}{\partial}_+ ,
\stackrel{\leftarrow}{\partial}_+) \,\lambda(0)\right] \, ,
$$
which possesses the same anomalous dimension as the operator (\ref{O2-N=1}).

We remind that the above expressions are valid in the axial gauge $(n\cdot
A(x))=A_+(x)=0$. To write down the same expression in a covariant form, it is
sufficient to substitute the light-cone by covariant derivatives $\partial_+ =
(n\cdot D)$, the antiholomorphic gauge field by the field strength tensors,
$\partial_+ \bar A = ( F_{+ x}^{\ \perp} - i {F}_{+ y}^{\ \perp}) /\sqrt{2}$,
and the Grassmann fermion $\lambda$ by the two-component Weyl spinor (see
Appendix A of Ref.\ \cite{BelDerKorMan04}).

\subsubsection*{$\mathcal{N}=2$ theory}

The light-cone superfield $\Psi(Z)$, Eq.~\re{M=2-field}, comprises the scalar
field $\phi$, the gauge field $\partial_+\bar A$ of helicity $-1$ and the gaugino
field $\lambda^A$ (with ${\scriptstyle A}=1,2$) of helicity $-1/2$. It carries
the superconformal spin $j=1/2$ and belongs to the irreducible $SL(2|2)$
representation, $\Psi(Z)\in \mathbb{V}_{1/2}$.

As before, to construct the Wilson operators in the $\Psi\Psi-$sector, one
identifies the lowest weights in the tensor product $\mathbb{V}_{1/2}\otimes
\mathbb{V}_{1/2}$, Eq.~\re{weights}. They are
\ba
\Psi_0^{(0)}(Z_1,Z_2)
\!\!\!&=&\!\!\! 1
\,,
\nonumber \\[2mm]
\Psi_1^{(0)}(Z_1,Z_2)
\!\!\!&=&\!\!\!
(\theta_1-\theta_2)^A
\,,
\label{Psi-N2} \\[2mm]
\Psi_n^{(0)}(Z_1,Z_2)
\!\!\!&=&\!\!\!
\varepsilon_{AB}(\theta_1-\theta_2)^A(\theta_1-\theta_2)^B
(z_1-z_2)^{n-2}
\,, \nonumber
\ea
where $n\ge 2$ and $Z=(z,\,\theta^A)$ (with ${\scriptstyle A}=1,2$). To determine
the corresponding superconformal polynomials one matches these expressions into
\re{Psi-ansatz} and \re{P-ansatz} and obtains
\ba
P_0^{(0)}( {X}_1; {X}_2)
\!\!\!&=&\!\!\! 1\,,
\nonumber \\[3mm]
P_1^{(0)}( {X}_1; {X}_2)
\!\!\!&=&\!\!\! (\vartheta_1-\vartheta_2)^A
\label{N2Pncc} \\[2mm]
P_n^{(0)}( {X}_1; {X}_2)
\!\!\!&=&\!\!\!
c_n (x_1+x_2)^{n-2}\left\{ -\frac{2n}{n-1}
\,\mathrm{P}_{n-2}^{(1,1)}\left(\frac{x_2-x_1}{x_2+x_1}\right)(\vartheta_1\cdot
\vartheta_2)\right.
\nonumber \\
&& +\left.
\mathrm{P}_{n-2}^{(2,0)}\left(\frac{x_2-x_1}{x_2+x_1}\right)(\vartheta_1\cdot
\vartheta_1)+\mathrm{P}_{n-2}^{(0,2)}\left(\frac{x_2-x_1}{x_2+x_1}\right)(\vartheta_2\cdot
\vartheta_2) \right\}\,,\nonumber
\ea
where $c_n=(-1)^n/n!$, $ {X}=(x,\vartheta_A)$ and the notation was introduced for
$(\vartheta_k\cdot \vartheta_n) \equiv \varepsilon^{AB} \vartheta_{k, A}
\vartheta_{n, B}$. Then, one substitutes \re{N2Pncc} into \re{O-N=1}, takes into
account that $ \Psi(Z)=i \phi(x) - \varepsilon_{AB} \theta^A \lambda^B (x) +
\frac12 \varepsilon_{AB} \theta^A\theta^B \partial_+
\bar A (x)
$ and obtains the following expressions:
\begin{itemize}
\item For $n=0$, the twist-two operator is built from the complex scalar
\be\label{phiphi}
\mathcal{O}_0^{(0)}(0) = \tr \left[\Psi(0)\Psi(0) \right] = -\tr
\left[\phi(0)\phi(0) \right]\,.
\ee
\item For $n=1$, the operator vanishes
\be
\mathcal{O}_1^{(0)}(0) = \tr
\left[\partial_{\theta^A}\Psi(0)\Psi(0)-\Psi(0)\partial_{\theta^A}\Psi(0) \right]
= 0\,.
\ee
\item For $n\ge 2$, the twist-two operators take the form (up to an overall
normalization factor)
\ba\label{O3-N=2}
\mathcal{O}_n^{(0)}(0) &=& \tr\left\{ -
\frac{in}{(n-1)}\varepsilon_{AB}\lambda^A(0)\,
\mathbb{P}^{(1,1)}_{n-2}(\stackrel{\rightarrow}{\partial}_+
,
\stackrel{\leftarrow}{\partial}_+)\, \lambda^B (0) \right.
\\
&& \left.+\phi(0)\, \mathbb{P}^{(0,2)}_{n-2}(\stackrel{\rightarrow}{\partial}_+ ,
\stackrel{\leftarrow}{\partial}_+)\, \partial_+ \bar A(0)+\partial_+ \bar A(0)\,
\mathbb{P}^{(2,0)}_{n-2}(\stackrel{\rightarrow}{\partial}_+ ,
\stackrel{\leftarrow}{\partial}_+)\,\phi(0) \right\}\,. \nonumber
\ea
\end{itemize}
The operators $\mathcal{O}_n^{(0)}(0)$ carry the superconformal spin $j = 2j_\Psi
+ n = 1+n$ and their one-loop anomalous dimension equals $E_{\Psi\Psi}(n+1)=2
\left[ \psi(n+1)-\psi(1)\right]$, Eqs.~\re{gg} and \re{H-PsiPsi-J}. For $n=0$,
the operator \re{phiphi} has a vanishing anomalous dimension and this value is
protected to all orders by supersymmetry~\cite{DiVTan04}. In the forward limit,
for $\stackrel{\rightarrow}{\partial}_+ + \stackrel{\leftarrow}{\partial}_+=0$,
the operator \re{O3-N=2} is given by
\be
\mathcal{O}_n^{(0)}(0) \,\, \stackrel{\rm fw}{\sim } \,\, \sigma_{n-2} \tr\left\{
- \frac{in}{(n-1)}\varepsilon_{AB}\lambda^A(0)\,{\partial}_+^{n-2}\,\lambda^B (0)
+ 2 \phi(0)\,  \partial_+^{n-1} \bar A(0)  \right\}
\, ,
\ee
which coincides with the results of Ref.~\cite{BelDerKorMan04}. Other components
of the supermuliplet are found from descendants of the lowest weights
(\ref{N2Pncc}) in the same vein as we explained above in $\mathcal{N} = 1$ SYM
theory.

\subsection{Superconformal operators in $\Phi\Psi-$ and $\Phi\Phi-$ sectors}

Let us extend the analysis to twist-two operators in the $\Psi\Phi-$ and
$\Phi\Phi-$sectors. The important difference between the light-cone superfields
$\Psi(Z)$ and $\Phi(Z)$, Eqs.~\re{M=0-field} -- \re{M=4-field}, is that the
latter involves {\it nonlocal} field operators, $\partial_+^{-1} A(x)$ and
$\partial_+^{-1} \lambda(x)$. As a consequence, the superfield $\Phi(Z)$
carries a {\it negative} superconformal spin $j_\Phi=-1/2$, independent on the
number of supercharges $\mathcal{N}$, and the corresponding $SL(2|\mathcal{N})$
representation $\mathbb{V}_{j_\Phi}$ is reducible.

It is easy to see that some of the formulae obtained in Sect.~3 are not well
defined for negative spin $j$. For instance, integrals entering \re{PPsi} and
\re{PPsi-S} are divergent for $j=-1/2$. Still, as we shall argue below, the
relations between the superconformal polynomials $P_n^{(0)}(X_1,X_2)$ and the
lowest weights $\Psi_n^{(0)}(Z_1,Z_2)$, Eqs.~\re{SL2N-F} and \re{PPsi-S}, are
still at work when analytically continued to $j_{1,2}=-1/2$.

Let us start with the $\mathcal{N}=0$ case and examine the Taylor expansion of
the field $\Phi(z n_\mu)$, Eq.~\re{M=0-field}, around the origin
\be\label{Phi-exp}
{\Phi} (z)=\left(\partial_+^{-1} A(0)+ z A(0)\right) + \sum_{k=2}^\infty \frac{
z^{k}}{k!}\,\partial_+^{k-1}A(0)\,.
\ee
Here the first two terms involve the operators $\partial_+^{-1}A(0)$ and $A(0)$.
Their appearance is an artefact of the light-cone formulation of the
$\mathcal{N}=0$ theory and in the covariant formulation they correspond to
nonlocal, ``spurious'' gauge field operators.

The field ${\Phi} (zn_\mu)$ defines a representation of the conformal $SL(2)$
group of spin $j=-1/2$. As before, it is spanned by the coefficient functions
$\mathbb{V}_{-1/2}=\{1,z,z^2,\ldots\}$ entering the Taylor expansion
\re{Phi-exp} and the $SL(2)$ generators are given by \re{SS} with $j=-1/2$. The
coefficient functions $\mathbb{V}_{\rm sp} = \{1,z\}$ accompanying the spurious
operators in \re{Phi-exp} define the $SL(2)$ invariant two-dimensional
subspace. For ``physical'', Wilson operators the corresponding coefficient
functions belong to the quotient of two spaces, $\mathbb{V}_{\rm
phys}=\mathbb{V}_{-1/2}/\mathbb{V}_{\rm sp}$.

We remind that for positive conformal spins $j$ the vectors $z_k\in \mathbb{V}_{j}$
have the $SL(2)$ norm $\vev{z^k|z^n} \sim \delta_{kn}/{\Gamma(k+2j)}$, Eq.~\re{SL2-norm}.
Analytically continuing this relation to $j=-1/2$ one finds that the vectors
belonging to $\mathbb{V}_{\rm sp} = \{1,z\}$ have a zero norm and, in addition,
they are orthogonal to the vectors from $\mathbb{V}_{\rm phys}$. Therefore, for
$j=-1/2$ the $SL(2)$ scalar product \re{sc} projects the field $\Phi(z n_\mu)$,
Eq.~\re{Phi-exp}, onto the subspace of physical operators
\be\label{dec}
\vev{z^k|\Phi(z)} =
\left\{\begin{array}{ll}
0\,, & k=0,1 \\
\frac1{(k-2)!}{\partial_+^{k-1}A(0)}\,, & k\ge 2 \\
\end{array}
\right.
\ee
Let us now consider the product of two $\mathcal{N}=0$ fields
$\tr[\Psi(z_1)\Phi(z_2)]$ and expand it in powers of $z_1$ and $z_2$ over the
set of multiplicatively renormalized operators. Depending on whether these
operators involve spurious gauge field operators, they can be split into two
groups\\[0.0mm]
\be\label{spurT}
\tr\left[\Psi(z_1)\Phi(z_2)\right]=\sum_{n,\ell} \Psi_n^{(\ell)}(z_1,z_2)\,
\mathcal{O}_n^{(\ell)}(0)~+~ \sum_{n,\ell} \widetilde\Psi_n^{(\ell)}(z_1,z_2)\,
\widetilde{\mathcal{O}}_n^{(\ell)}(0)\,,
\ee\\[-2mm]
where $\mathcal{O}_n^{(\ell)}(0)$ denote ``physical'' twist-two operators. The
``spurious'' operators $\widetilde{\mathcal{O}}_n^{(\ell)}(0)$ involve
$\partial_+^{-1}A(0)$ and $A(0)$ and, as a consequence, $\widetilde
\Psi_n^{(\ell)}(z_1,z_2)$ is linear in $z_2$. By virtue of \re{dec}, $\widetilde
\Psi_n^{(\ell)}(z_1,z_2)$ has zero projection onto all states in
$\mathbb{V}_{3/2}\otimes \mathbb{V}_{-1/2}$. Therefore, taking the scalar product
of both sides of \re{spurT} with the lowest weights $\Psi_n^{(0)}(Z_1,Z_2)\in
\mathbb{V}_{3/2}\otimes \mathbb{V}_{-1/2}$ one finds that ``spurious'' operators
do not contribute and only ``physical'' operators survive.

Indeed, let us examine the expression for the $SL(2)$ conformal operator
\re{gen-def} for $j_1=-1/2$ and $j_2=3/2$ corresponding to the conformal spins
of $\Phi-$ and $\Psi-$fields, Eq.~\re{M=0-field}. Using the properties of the
Jacobi polynomials (see Eqs.~\re{F1} and \re{F2}) one obtains (for $n\ge 2$)
\ba\label{anom-N=0}
{\mathcal O}_{n}^{-1/2,3/2}(0) &=& \tr\left[ \Phi(0)\,
\mathbb{P}^{(-2,2)}_{n}
\left({\stackrel{\rightarrow}{\partial}_+, \stackrel{\leftarrow}{\partial}_+}
\right) \Psi(0)\right]
\\
&\sim& \tr\left[\partial_+^{2} \Phi(0)\,
\mathbb{P}^{(2,2)}_{n-2}
\left({\stackrel{\rightarrow}{\partial}_+, \stackrel{\leftarrow}{\partial}_+}
\right) \Psi(0)\right]={\mathcal O}_{n-2}^{3/2,3/2}(0)\,, \nonumber
\ea
where $\partial_+^{2} \Phi(0)=\partial_+ A(0)$, $\Psi=-\partial_+ \bar A(0)$ and
the $\mathbb{P}-$operator was defined in \re{P-op}. Eq.~\re{anom-N=0} defines the
$SL(2)$ conformal operator in the $\mathcal{N}=0$ theory in the
$\Phi\Psi-$sector. This operator has the $SL(2)$ conformal spin
$j=j_\Psi+j_\Phi+n=1+n$ and its one-loop anomalous dimension is given by
$E_{\Psi\Phi}(n+1)$, Eqs.~\re{gg} and \re{H-PhiPsi-J}, evaluated for
$\mathcal{N}=0$. In a similar manner, for $j_1=j_2=-1/2$ one finds the conformal
operators in the $\Phi\Phi-$sector (for $n\ge 4$)
\ba\label{anom1-N=0}
{\mathcal O}_{n}^{-1/2,-1/2}(0) &=& \tr\left[ \Phi(0)\, \mathbb{P}^{(-2,-2)}_{n}
\left({\stackrel{\rightarrow}{\partial}_+, \stackrel{\leftarrow}{\partial}_+}
\right) \Phi(0)\right]
\\
&=& \tr\left[\partial_+^{2} \Phi(0)\,
\mathbb{P}^{(2,2)}_{n-2}
\left({\stackrel{\rightarrow}{\partial}_+, \stackrel{\leftarrow}{\partial}_+}
\right) \partial_+^{2} \Phi(0)\right]={\mathcal O}_{n-4}^{3/2,3/2}(0)\,,
\nonumber
\ea
where $\partial_+^{2} \Phi(0)=\partial_+ A(0)$. This operator has the conformal
spin $j=2j_\Phi+n=-1+n$ and its one-loop anomalous dimension is given by
$E_{\Phi\Phi}(n-1)$, Eqs.~\re{gg} and \re{H-PsiPsi-J}.%
\footnote{By virtue of the relation $\partial_+^{2} \Phi(0) = -\widebar{\Psi(0)}
=\partial_+ A(0)$ (see Eq.~\re{M=0-field}), the twist-two operators in the
$\Phi\Phi-$ and $\Psi\Psi-$sectors are complex conjugated to each other.}

Let us extend the analysis to $\mathcal{N}\ge 1$. Examining the expansion of the
superfield $\Phi(z n_\mu,\theta^A)$, Eq.~\re{M=1-field} -- \re{M=4-field}, around
the origin $z=\theta^A=0$ one observes that the coefficient functions
accompanying nonlocal, ``spurious'' field components define the
$SL(2|\mathcal{N})$ invariant subspace $\mathbb{V}_{\rm sp} = \{1,z,\theta^A,
z\theta^A\}$ (with ${\scriptstyle A}=1,\ldots,\mathcal{N}$). Applying \re{norm-N}
for $j=-1/2$ one finds that the states belonging to $\mathbb{V}_{\rm sp}$ have a
zero norm and they are orthogonal to all states in $\mathbb{V}_{j_\Phi}$.
Therefore, the scalar product of $\tr\left[\Phi(Z_1)\Psi(Z_2)\right]$ and
$\tr\left[\Phi(Z_1)\Phi(Z_2)\right]$ with the $SL(2|\mathcal{N})$ lowest weights
$\Psi_n^{(0)}(Z_1,Z_2)$ eliminates the contribution of operators involving
nonlocal field components and only retains Wilson operators. This allows one to
use a general expression for the superconformal polynomials, Eqs.~\re{SL2N-F} and
\re{P-ansatz}, and analytically continue it to the values of superconformal spins
$j_\Phi=-1/2$ and $j_\Psi=(3-\mathcal{N})/2$ corresponding to the $\Phi-$ and
$\Psi-$superfields. Notice that the lowest weights, Eqs.~\re{weights} and
\re{Psi-ansatz}, do not depend on the superconformal spins of the superfields.
This dependence enters into \re{P-ansatz} only through the coefficients
${c_N^{(2j_1+k_1,2j_2+k_2)}}$ and indices of the Jacobi polynomials.

\subsubsection*{$\mathcal{N}=1$ theory}

Let us first consider the twist-two operators in the $\Phi\Phi-$sector for
$\mathcal{N}=1$ and $\mathcal{N}=2$. According to \re{M=1-field} and
\re{M=2-field}, the light-cone superfields $\Phi(Z)$ and $\Psi(Z)$,
Eqs.~\re{M=0-field} -- \re{M=4-field} involve two different sets of the
fundamental fields which are mutually conjugated to each other. This suggests
that the same relation should hold between the superconformal operators in the
$\Phi\Phi-$ and $\Psi\Psi-$sectors. Indeed, substituting $j_1=j_2=-1/2$ into
\re{P-ansatz} and going over through the calculation of the expansion
coefficients $c_N^{(2j_1+k_1,2j_2+k_2)}$, Eq.~\re{cc}, one finds that for
$\mathcal{N}=1$ and $\mathcal{N}=2$ the superconformal operators in the
$\Phi\Phi-$sector are given by the same expressions as before, Eqs.~\re{O1-N=1},
\re{O2-N=1} and Eqs.~\re{phiphi} -- \re{O3-N=2}, respectively, provided that the
fields $\partial_+\bar A$, $\lambda^A$, $\phi$ are replaced by the corresponding
conjugated fields $\partial_+ A$, $\bar \lambda_A$, $\bar\phi$. Obviously, this
substitution does not affect anomalous dimensions of Wilson operators.

In the $\Phi\Psi-$sector, the superconformal operators are defined as
\be
\mathcal{O}_n^{(0)}(0) =P_{n}^{(0)}( \partial_{{Z}_1};\partial_{{Z}_2}) \tr
\left[ \Phi(Z_1) \Psi(Z_2)\right]\bigg|_{Z_1=Z_2=0}\,,
\ee
with the polynomials given by \re{P-ansatz}  for $j_1=-1/2$ and
$j_2=(3-\mathcal{N})/2$ and the lowest weights \re{Psi-ansatz} of the form
\re{weights}.

For $\mathcal{N}=1$, the superconformal polynomials corresponding to the lowest
weights \re{Psi-N1} are $P_{k}^{(0)}( {X}_1; {X}_2) = 0$ for $k=0,1$ and
\be
\label{deg-N1}
P_{n}^{(0)}( {X}_1; {X}_2) =
(x_1+x_2)^{n-1}\left\{b_{n}^{(1)}\,
\mathrm{P}_{n-1}^{(-1,1)}\left(\frac{x_2-x_1}{x_2+x_1}\right)
\vartheta_1
-
b_{n}^{(2)}\,\mathrm{P}_{n-1}^{(-2,2)}\left(\frac{x_2-x_1}{x_2+x_1}\right)
\vartheta_2
\right\},
\ee
with $b_n^{(l)} = (-1)^{n-1} \Gamma(n)/[\Gamma(n+l)\Gamma(n-l)]$. Applying the
identities \re{F1} and \re{F2} one obtains%
\footnote{Here and in what follows it is tacitly assumed that the Jacobi
polynomials $\mathrm{P}_{n}^{(a,b)}$ vanish for negative $n$.}
\be
P_{n}^{(0)}( {X}_1; {X}_2) =- b_{n}^{(0)}{(x_1+x_2)^{n-1}}
\left[\mathrm{P}_{n-2}^{(1,1)}\left(\frac{x_2-x_1}{x_2+x_1}\right)\frac{x_1
\vartheta_1}{x_1+x_2}+\mathrm{P}_{n-3}^{(2,2)}\left(\frac{x_2-x_1}{x_2+x_1}\right)
\frac{x_1^2 \vartheta_2}{(x_1+x_2)^2}  \right]\,,
\ee
for $n\ge 2$. It follows from \re{M=1-field} that
\ba
&&
\partial_{z_1}\partial_{\theta_1} \Phi(Z_1) \Psi(Z_2) \big|_{\theta_{1,2}=0}
= -\bar \lambda(z_1) \, \lambda(z_2) \,,\qquad \nonumber
\\
&&
\partial_{z_1}^2\partial_{\theta_2} \Phi(Z_1) \Psi(Z_2) \big|_{\theta_{1,2}=0} =
\partial_+ A(z_1) \, \partial_+\bar A(z_2)\,.
\ea
Therefore, the twist-two operators defined by the superconformal polynomials
\re{deg-N1} are
\be\label{OO-N=1}
\mathcal{O}_n^{(0)}(0) = \tr\left\{ \bar \lambda
(0)\,\mathbb{P}^{(1,1)}_{n-2}(\stackrel{\rightarrow}{\partial}_+ ,
\stackrel{\leftarrow}{\partial}_+)\, \lambda  (0) - \partial_+ A(0)\,
\mathbb{P}^{(2,2)}_{n-3}(\stackrel{\rightarrow}{\partial}_+ ,
\stackrel{\leftarrow}{\partial}_+)\, \partial_+ \bar A(0)  \right\}
\ee
with $n\ge 2$. They carry the superconformal spin $j=j_\Phi+j_\Psi+n=n+1/2$ and
their one-loop anomalous dimension equals $E_{\Phi\Psi} (n+1/2)$, Eqs.~\re{gg}
and \re{H-PhiPsi-J} for $\mathcal{N}=1$. In the forward limit, for
$\stackrel{\rightarrow}{\partial}_+ + \stackrel{\leftarrow}{\partial}_+=0$, the
operator \re{OO-N=1} is given by
\be
\mathcal{O}_n^{(0)}(0) \,\, \stackrel{\rm fw}{\sim } \,\,
\tr\left\{ \bar \lambda(0)\,\partial_+^{n-2}\, \lambda  (0)
-
\frac{n-2}{n+1} \partial_+ A(0)\, \partial_+^{n-2} \bar A(0) \right\}
\,,
\ee
which is the results obtained in Ref.~\cite{BelDerKorMan04}.

\subsubsection*{$\mathcal{N}=2$ theory}

For $\mathcal{N}=2$ the lowest weight are given by \re{Psi-N2} and the conformal
spins of the superfields are $j_\Phi=-1/2$ and $j_\Psi=1/2$. One applies
\re{P-ansatz} and calculates the corresponding superconformal polynomials as
$P_{k}^{(0)}( {X}_1; {X}_2)=0$ for $k=0,1$ and
\ba
P_n^{(0)}( {X}_1; {X}_2)
\!\!\!&=&\!\!\!
(x_1+x_2)^{n-2}\left\{b_{n-1}^{(2)}\,
\mathrm{P}_{n-2}^{(-2,2)}\left(\frac{x_2-x_1}{x_2+x_1}\right)(\vartheta_2\cdot
\vartheta_2)\right.
\\
&& \left.-2 b_{n-1}^{(1)}
\,\mathrm{P}_{n-2}^{(-1,1)}\left(\frac{x_2-x_1}{x_2+x_1}\right)(\vartheta_1\cdot
\vartheta_2) +
b_{n-1}^{(0)}\,\mathrm{P}_{n-2}^{(0,0)}\left(\frac{x_2-x_1}{x_2+x_1}\right)
(\vartheta_1 \cdot \vartheta_1) \right\}
\, , \nonumber
\ea
for $n\ge 2$ and the $b-$coefficients defined in \re{deg-N1}. Simplification of
the Jacobi polynomials with a help of \re{F1} and \re{F2} yields
\ba
P_n^{(0)}( {X}_1; {X}_2)
\!\!\!&=&\!\!\!
b_{n-1}^{(0)}(x_1+x_2)^{n-2}
\left\{\mathrm{P}_{n-4}^{(2,2)}
\left(\frac{x_2-x_1}{x_2+x_1}\right)
\frac{(\vartheta_2\cdot \vartheta_2)x_1^2}{(x_1+x_2)^{2}} \right.
\nonumber \\
&&+\left.
2\mathrm{P}_{n-3}^{(1,1)}\left(\frac{x_2-x_1}{x_2+x_1}\right)\frac{(\vartheta_1\cdot
\vartheta_2)x_1}{x_1+x_2}+
\mathrm{P}_{n-2}^{(0,0)}\left(\frac{x_2-x_1}{x_2+x_1}\right)(\vartheta_1\cdot
\vartheta_1) \right\} \,.
\ea
Its substitution into \re{SuperConfOper} leads to the following expression for
the twist-two operator (up to an overall normalization factor)
\ba
\mathcal{O}_n^{(0)}(0) \!\!\!&=&\!\!\! \tr\bigg\{ \bar \phi
(0)\,\mathbb{P}^{(0,0)}_{n-2}(\stackrel{\rightarrow}{\partial}_+ ,
\stackrel{\leftarrow}{\partial}_+)\, \phi  (0) + \bar \lambda_A
(0)\,\mathbb{P}^{(1,1)}_{n-3}(\stackrel{\rightarrow}{\partial}_+ ,
\stackrel{\leftarrow}{\partial}_+)\, \lambda^A  (0)
\nonumber \\
&& \qquad  - \partial_+ A(0)\,
\mathbb{P}^{(2,2)}_{n-4}(\stackrel{\rightarrow}{\partial}_+ ,
\stackrel{\leftarrow}{\partial}_+)\, \partial_+ \bar A(0)
\bigg\}\,.
\label{OO-N=2}
\ea
This operator has the superconformal spin $j=j_\Phi+j_\Psi+n=n$ and its one-loop
anomalous dimension is given by $E_{\Phi\Psi}(n)$, Eqs.~\re{gg} and
\re{H-PhiPsi-J} for $\mathcal{N}=2$.  In the forward limit, for
$\stackrel{\rightarrow}{\partial}_+ + \stackrel{\leftarrow}{\partial}_+=0$, the
operator \re{OO-N=2} is given by
\be
\mathcal{O}_n^{(0)}(0) \,\, \stackrel{\rm fw}{\sim } \,\, \tr\bigg\{
\partial_+A(0)\,\partial_+^{n-3} \bar A(0) -
\frac{n}{n-3}\bar \lambda_A (0)\,\partial_+^{n-3}\, \lambda^A  (0)-
\frac{n(n-1)}{(n-2)(n-3)}\bar \phi (0)\,\partial_+^{n-2}\, \phi  (0) \bigg\}\,,
\ee
which agrees with the expression obtained in Ref.~\cite{BelDerKorMan04}.

\subsubsection*{$\mathcal{N}=4$ theory}

A unique feature of the $\mathcal{N}=4$ theory is that all twist-two operators
belong to the $\Phi\Phi-$sector. The superconformal spin of the $\Phi-$superfield
equals $j_\Phi=-1/2$ and the lowest weights in the tensor product of two
$SL(2|4)$ modules are given for $\mathcal{N}=4$ by \re{weights}
\ba
\Psi_0^{(0)}(Z_1,Z_2)
\!\!\!&=&\!\!\!
1\,,
\nonumber \\[2mm]
\Psi_{n<4}^{(0)}(Z_1,Z_2)
\!\!\!&=&\!\!\!
\varepsilon_{A_1A_2A_3A_4} \prod_{k=1}^n (\theta_1-\theta_2)^{A_k}
\,,
\label{Psi-N4} \\[-1mm]
\Psi_{n\ge 4}^{(0)}(Z_1,Z_2)
\!\!\!&=&\!\!\!
(z_1-z_2)^{n-4} \varepsilon_{A_1A_2A_3A_4}
\prod_{k=1}^4 (\theta_1-\theta_2)^{A_k}
\,. \nonumber
\ea
Let us match these expressions into \re{Psi-ansatz} and \re{P-ansatz}. For the
lowest weights $\Psi_{n}^{(0)}(Z_1,Z_2)$ with $0\le n < 4$ the expansion
coefficients entering \re{P-ansatz} take the form
$c_0^{(-1+k_1,-1+k_2)}=1/[\Gamma(k_1-1)\Gamma(k_2-1)]$ and are different from
zero provided that $k_1, k_2 \ge 2$. Since $k_1+k_2=n$, one concludes that
$c_0^{(-1+k_1,-1+k_2)}=0$  for $n \ge 3$ leading to
\be
\label{zero-N=4}
P_{n}^{(0)}(X_1,X_2) =0\,, \qquad (0 \le n<  4)
\ee
For $n \ge 4$, the superconformal polynomial \re{P-ansatz} corresponding to the
lowest weight $\Psi_{n\ge 4}^{(0)}(Z_1,Z_2)$ takes the form
\ba\label{P-N=4}
P_{n}^{(0)}(X_1,X_2) \!\!\!&=&\!\!\! 4! \sum_{k= n_{\rm min} }^{4-n_{\rm
min}}(-1)^k b_{n-3}^{(k-2)}\,(x_1+x_2)^{n-4}\,
\mathrm{P}_{n-4}^{(k-2,2-k)}\left(\frac{x_2-x_1}{x_2+x_1}\right)
\\
&& \times\, \varepsilon^{A_1 A_2 A_3 A_4} \prod_{m_1=1}^{k} \vartheta_{1,A_{m_1}}
\prod_{m_2=k+1}^{4} \vartheta_{2,A_{m_2}}
\,, \nonumber
\ea
where $n_{\rm min} \equiv {\rm max}(6-n,0)$, $X_k=(x_k,\vartheta_{k,A})$ and
the $b-$coefficients were defined in \re{deg-N1}. Here, the ordering of
$\vartheta-$variables is such that $\prod_{m=1}^{\ell}
\vartheta_{A_m}=\vartheta_{A_1}\ldots\vartheta_{A_\ell}$.

For $n=4$, the relation \re{P-N=4} takes a simple form
\be\label{n=4-N=4}
P_{4}^{(0)}(X_1,X_2)= 4! \, b_{1}^{(0)}\,
\varepsilon^{A_1 A_2 A_3 A_4}
\vartheta_{1,A_{1}}\vartheta_{1,A_{2}}
\vartheta_{2,A_{3}}\vartheta_{2,A_{4}}
\, .
\ee
For $n \ge 5$, one of the indices of the Jacobi polynomial in \re{P-N=4} takes
negative values. Applying the identities \re{F1} and \re{F2} one obtains after
some algebra (for $\ell=k-2)$
\ba\label{Pol-N=4}
&&
P_{n}^{(0)}(X_1,X_2) = 4!\, b_{n-3}^{(0)}\, \sum_{\ell= 0 }^{2-n_{\rm min}}
\frac{\kappa_\ell}{(2+\ell)!(2-\ell)!}\,
(x_1+x_2)^{n-4-\ell}\,\mathrm{P}_{n-4-\ell}^{(\ell,\ell)}\left(\frac{x_2-x_1}{x_2+x_1}\right)
\\
&& \qquad \times\,
\varepsilon^{A_1 A_2 A_3 A_4}\left[ (-x_2)^\ell
\prod_{m_1=1}^{2+\ell} \vartheta_{1,A_{m_1}}\!\!\!\! \prod_{m_2=3+\ell}^{4}
\vartheta_{2,A_{m_2}} + x_1^\ell \prod_{m_1=1}^{2-\ell}
\vartheta_{1,A_{m_1}}\!\!\!\! \prod_{m_2=3-\ell}^{4} \vartheta_{2,A_{m_2}}
\right] \,, \nonumber
\ea
with $\kappa_{\ell=0}=1/2$ and $\kappa_{\ell\neq 0}=1$.

Let us translate these expressions into the superconformal operators
\be\label{Op-N=4}
\mathcal{O}_n^{(0)}(0) =P_{n}^{(0)}( \partial_{{Z}_1};\partial_{{Z}_2}) \tr
\left[ \Phi(Z_1) \Phi(Z_2)\right]\bigg|_{Z_1=Z_2=0}\,,
\ee
with the $\Phi-$superfield given by \re{M=4-field}. One finds from \re{zero-N=4}
that $\mathcal{O}_n^{(0)}(0)$ vanishes for $0\le n < 4$. For $n=4$, it follows
from \re{n=4-N=4} that (up to an overall normalization factor)
\be
\mathcal{O}_4^{(0)}(0) = \frac14 \varepsilon^{A_1 A_2 A_3 A_4}
\partial_{\theta_1^{A_{1}}}\partial_{\theta_1^{A_{2}}}
\partial_{\theta_2^{A_{3}}}\partial_{\theta_2^{A_{4}}}
\tr \left[\Phi(Z_1)\Phi(Z_2)\right]\big|_{Z_{1,2}=0}
= -\frac12 \tr \left[\bar\phi_{A_1A_2}(0) \phi^{A_1A_2}(0)\right]\,,
\ee
where $\phi^{A_1 A_2} = \frac12 \varepsilon^{A_1 A_2 A_3 A_4}\bar\phi_{A_3 A_4}$.
For $n \ge 5$, the substitution of \re{Pol-N=4} into \re{Op-N=4} yields
\ba
\mathcal{O}_n^{(0)}(0)
\!\!\!&=&\!\!\!
\sigma_n \tr\bigg\{2\, \bar \lambda_A
(0)\,\mathbb{P}^{(1,1)}_{n-5}(\stackrel{\rightarrow}{\partial}_+ ,
\stackrel{\leftarrow}{\partial}_+)\, \lambda^A  (0)
\nonumber \\
&& -2\, \partial_+ A(0)\,
\mathbb{P}^{(2,2)}_{n-6}(\stackrel{\rightarrow}{\partial}_+ ,
\stackrel{\leftarrow}{\partial}_+)\, \partial_+ \bar A(0)
-
\frac12\bar \phi_{AB} (0)\,\mathbb{P}^{(0,0)}_{n-4}(\stackrel{\rightarrow}{\partial}_+
, \stackrel{\leftarrow}{\partial}_+)\, \phi^{AB}  (0)  \bigg\}
\,,
\label{OO-N=4}
\ea
with $\sigma_n=[1+(-1)^n]/2$. It reproduces the operator $\mathcal{S}^3_{n + 5}$
from Ref.\ \cite{BelDerKorMan03}. This operator carries the $SL(2|4)$ superconformal
spin $j=2j_\Phi+n=n-1$ and its one-loop anomalous dimension is given by
$E_{\Phi\Phi}(n-1)$, Eqs.~\re{gg} and \re{H-PsiPsi-J}. In the forward limit,
i.e., for $\stackrel{\rightarrow}{\partial}_+ + \stackrel{\leftarrow}{\partial}_+
=0$, the operator \re{OO-N=4} is reduces to
\ba
\mathcal{O}_n^{(0)}(0) \,\,
\stackrel{\rm fw}{\sim }
\,\, \sigma_n\tr\bigg\{
\partial_+A(0)\,\partial_+^{n-5} \bar A(0)
\!\!\!&-&\!\!\!
{\frac{n-2}{n-5}}\bar \lambda_A (0)\,\partial_+^{n-5}\, \lambda^A (0)
\nonumber\\
&-&\!\!\!
\frac{(n-2)(n-3)}{4(n-4)(n-5)}\bar \phi_{AB} (0)\,\partial_+^{n-4}\, \phi^{AB}
(0)
\bigg\}
\, .
\ea
Again it is in accordance with Refs.~\cite{KotLip02,BelDerKorMan04}.

We remind that the twist-two operators constructed in this section are the lowest
weights of the superconformal $SL(2|\mathcal{N})$ group. The remaining twist-two
operators belonging to the same supermultiplet are defined by the ``descendant''
superconformal polynomials $P_n^{(\ell)}(X_1,X_2)$ with $\ell \ge 1$. These
polynomials are obtained from the lowest weight $P_n^{(0)}(X_1,X_2)$ by applying
the raising operators \re{hat-ops}.

\section{Conclusions}

In the present work we have formulated an efficient framework for constructing
supermultiplets of conformal operators in supersymmetric Yang-Mills theories
with an arbitrary number of supercharges. These operators belong to irreducible
representations of the ``collinear'' $SL(2|\mathcal{N})$ group and their mixing
with other operators is protected to one-loop order by superconformal symmetry.
The central point in our analysis was representation for the superconformal
operators in terms of the superfields within the light-cone formalism. The
latter turns out to be advantageous as compared with the conventional covariant
formulation as it naturally accommodates supersymmetric models with an
arbitrary number of supersymmetries $0 \leq \mathcal{N} \leq 4$ and treats them
in the same fashion.

Our approach is based on a map of local Wilson operators into polynomials
defining the lowest weights in the tensor product of two representations of the
``collinear'' $SL(2| \mathcal{N})$ supergroup. As an outcome of the present
analysis we have found a concise expression for all superconformal operators of
twist two in supersymmetric Yang-Mills theories with $0 \le \mathcal{N} \le 4$.
It has a remarkably simple form and exhibits universal features which are not
obvious or hidden if addressed by other means. When written in components, the
obtained expressions coincide with the known twist-two operators derived by the
conventional technique based on a brute force inspection of transformation
properties of local Wilson operators under the action of supersymmery generators
and closure of their algebra \cite{BukFroLipKur85,BelDerKorMan03}. These
operators form a basis of eigenstates of the one-loop dilation operators in the
underlying gauge theories and enter into the operator product expansion of
diverse correlation functions, see, e.g., \cite{DolOsb01,AruDolOsbSok03} and
references therein.

\section*{Acknowledgements}

This work was supported in part by the grant 03-01-00837 of the Russian
Foundation for Fundamental Research (A.M. and S.D.) and by the grant VH-NG-004 of
the Helmholtz Association (A.M.). One of us (G.K.) is grateful to V.~Braun for
useful discussions and hospitality at the Institute for Theoretical Physics,
University of Regensburg.

\appendix

\setcounter{section}{0} \setcounter{equation}{0}
\renewcommand{\theequation}{\Alph{section}.\arabic{equation}}

\section{Appendix A: Jacobi polynomials}

The $SL(2)$ conformal operator  ${\mathcal O}_{n}^{j_1,j_2}(0)$,
Eq.~\re{gen-def}, is expressed in terms of the Jacobi polynomial
$\mathrm{P}_n^{(2j_1-1,2j_2-1)}(x)$. This polynomial is defined as
\be\label{Jacobi-def}
\mathrm{P}_n^{(\alpha,\beta)}(x)=\frac{\Gamma(n+1+\beta)}{n!\,\Gamma(1+\beta)}
\left(\frac{x-1}2\right)^n {}_2F_1 \left({{-n,-n-\alpha}\atop
{\beta+1}}\bigg|\frac{x+1}{x-1}\right)
\ee
and has the following properties:
\begin{itemize}
\item parity
\be
\mathrm{P}_n^{(\alpha,\beta)}(x) =  (-1)^n \mathrm{P}_n^{(\beta,\alpha)}(-x)
\ee
\item asymptotic behaviour
\be\label{as}
\mathrm{P}_n^{(\alpha,\beta)}(x)={\frac {\Gamma  \left( 2n+\alpha+\beta+1\right)
}{\Gamma  \left( n+\alpha+\beta+1\right)  n! }} (x/2)^n + \mathcal{O}(x^{n-1})
\ee
\item relation to the Gegenbauer polynomial
\be\label{P=C}
\mathrm{P}_n^{(\alpha,\alpha)}(x) = \mathrm{C}_n^{\alpha+1/2}(x)
\frac{\Gamma(2\alpha+1)\Gamma(n+\alpha+1)}{\Gamma(n+2\alpha+1)\Gamma(\alpha+1)}
\ee
\end{itemize}
The $SL(2)$ conformal operators ${\mathcal O}_{n}^{j_1,j_2}(0)$, ${\mathcal
O}_{n+2j_1-1}^{1-j_1,j_2}(0)$ and ${\mathcal O}_{n+2j_1+2j_2-2}^{1-j_1,1-j_2}(0)$
carry the same conformal spin $j_1+j_2+n$ and are not independent. The relations
between them can be established with a help of two identities (with $\alpha,\beta
\ge 0$)
\ba
\label{F1}
 \mathrm{P}_{n}^{(-\alpha,\beta)}(x) &=& \left(\frac{x-1}2\right)^\alpha
\mathrm{P}_{n-\alpha}^{(\alpha,\beta)}(x)
\frac{\Gamma(n+1-\alpha)\Gamma(n+1+\beta)}{\Gamma(n+1-\alpha+\beta)\Gamma(n+1)}\,,
\\
\label{F2}
\mathrm{P}_{n}^{(-\alpha,-\beta)}(x) &=& \left(\frac{x-1}2\right)^{\alpha}
\left(\frac{x+1}2\right)^{\beta}
\mathrm{P}_{n-\alpha-\beta}^{(\alpha,\beta)}(x)\,,
\ea
valid for $n-\alpha\ge 0$ and $n-\alpha-\beta\ge 0$, respectively.

\section{Appendix B: $SL(2|\mathcal{N})$ collinear group}

The superfield ${\Phi}_j(Z)\equiv {\Phi}_j(z,\theta^A)$ defines a representation
of the $SL(2|\mathcal{N})$ group labelled by its superconformal spin,
$\mathbb{V}_j$. The superconformal invariance implies that the evolution equation
\re{EQ} has to be invariant under the $SL(2|\mathcal{N})$ transformations of
superfields \re{local} generated by the operators \re{sl2}.

The scalar product on $\mathbb{V}_j$ is given by \re{ansatz-N} and
\re{measure-N}. It is invariant under the superconformal transformations $\Psi(Z)
\to \e^{i G} \Psi(Z)$
\be
\vev{\e^{i G} \Psi|\e^{i G}\Psi'} = \vev{\e^{-i G^\dagger}\e^{i G} \Psi|\Psi'}
=\vev{\Psi|\Psi'}\,,
\ee
provided that $G$, given by a linear combination of the $SL(2|\mathcal{N})$
generators \re{sl2}, is a self-adjoint operator, $G^\dagger = G$. In particular,
the operators ${L}^-$, ${L}^+$ and ${L}^0$ generate projective, $SU(1,1)$
transformations on the light-cone
\ba
\e^{i\alpha L^0}{\Phi}_j(Z) &=& (\e^{i \alpha })^j
\Phi_j(\e^{i\alpha}z,\e^{\frac{i}2\alpha} \theta^A)\,,
\nonumber \\
\e^{i\alpha L^+-i\bar\alpha L^-}{\Phi}_j(Z) &=&\frac{1}{(\bar b  z+\bar
a)^{2j}}\,{\Phi}_j\left(\frac{az+b}{\bar b  z+\bar a}, \frac{\theta^A}{\bar b
z+\bar a}\right)\,,
\label{SL2-subgroup}
\ea
with $a=\cosh |\alpha|$ and $b=i (\bar\alpha/\alpha)^{1/2}\sinh |\alpha|$. The
odd operators ${W}{}^{A,\pm}$ and ${V}^\pm_{A}$ generate conformal
transformations in the superspace and satisfy \re{conj-N}. For arbitrary
$\mathcal{N}$ these transformations have a rather complicated form whereas for
$\mathcal{N}=1$ they look as
\ba
\e^{\bar\varepsilon\, W^- + \varepsilon V^+}{\Phi}_j(Z) &=&{\Phi}_j\big( (z +
\bar\varepsilon\,\theta)\rho,(\theta + \varepsilon z)/\rho\big)\,,
\nonumber \\
\e^{\bar\varepsilon\, W^+ + \varepsilon V^-}{\Phi}_j(Z) &=&
\frac{\rho^{2j}}{(1-\bar\varepsilon\, \theta)^{2j}}{\Phi}_j\left(\frac{
z\,\rho}{1-\bar\varepsilon\, \theta}, \theta + \varepsilon\right)\,,
\label{V-plus}
\ea
where $\rho =1+\bar\varepsilon\varepsilon/2$.

The superfield ${\Phi}_j(Z)$ admits the following integral representation
\be\label{id}
\Phi_j(Z)=\Gamma(2j) \int [\mathcal{D}W]_j\,{(1-Z\cdot \bar W)^{-2j}}\,
{\Phi_j(W)}\,,
\ee
where $Z=(z,\theta^A)$, $W=(w,\vartheta^A)$ and $W\cdot \bar Z = w \bar z +
\vartheta^A\,\bar\theta_A$. To verify \re{id}, one substitutes $\Phi_j(W)$ by
its Taylor expansion around the origin $\Phi_j(W)\sim w^n \vartheta^{A_1}\ldots
\vartheta^{A_L}$ and performs integration with the help of \re{norm-N}.
Applying \re{id}, the superconformal operator \re{SuperConfOper}  can be
written as
\be\label{id1}
\mathcal{O}_n^{(0)}(0) = P_n^{(0)}\left(\partial_{Z_1},\partial_{Z_2}\right) \int
\prod_{k=1}^2\, [\mathcal{D} W_k]_{j_k} \Gamma(2j_k) (1-Z_k\cdot \bar
W_k)^{-2j_k} \mathbb{O}(W_1,W_2)\bigg|_{Z_{1,2}=0}\,.
\ee
Matching this identity into \re{N-v}, one arrives at the relation \re{PPsi-S}
which is valid for arbitrary $\mathcal{N}$. For $\mathcal{N}=0$, it leads to
\re{PPsi}.

A general expression for the superconformal polynomial $P(X_1,X_2)$ is given by
\re{P-ansatz}. To obtain this expression, one substitutes \re{Psi-ansatz} into
\re{SL2N-F} and applies \re{dec1}
\ba
&& P(X_1,X_2) = \sum_{k_1,k_2=0}^\mathcal{N}
\frac1{(\mathcal{N}-k_1)!(\mathcal{N}-k_2)!}\int [\mathcal{D} z_1]_{j+k_1/2}\int
[\mathcal{D} z_2]_{j+k_2/2}\, \bar{z}_{12}^n \,\textrm{e}^{\,z_1x_1+z_2x_2}
\label{new} \\
& & \times \int \prod_{A=1}^\mathcal{N} \left(d\bar\theta_{1,A}d\theta_1^A
\right) \int \prod_{A=1}^\mathcal{N} \left(d\bar\theta_{2,A}d\theta_2^A
\right)\,
 {(\bar\theta_1\cdot \theta_1)^{\mathcal{N}-k_1}} {(\bar\theta_2\cdot
\theta_2)^{\mathcal{N}-k_2}}
 \, \widebar{\varphi_M(\theta_{12})}\,
\e^{\theta_1\cdot\vartheta_1+\theta_2\cdot\vartheta_2}. \nonumber
\ea
The $z-$integral is the same as in \re{SL2-Four}, and is given by the Jacobi
polynomial \re{Jacobi}. The $\theta-$integral can be easily evaluated with a help
of \re{theta-rules}.



\end{document}